\documentclass[fleqn,usenatbib]{mnras}

\usepackage{mathptmx}

\usepackage[T1]{fontenc}
\usepackage{ae,aecompl}

\usepackage{graphicx}	
\usepackage{amsmath}	
\usepackage{amssymb}	
\usepackage{natbib}
\usepackage{bm}			

\newcommand{\nab}{\bm{\nabla}}

\title[Fast-oscillation damping by convection]{Turbulent damping of fast tidal oscillations by three-dimensional Rayleigh–Bénard convection with a radiating free surface}

\author[C. Terquem et al.]{ Caroline Terquem$^{1,2}$\thanks{caroline.terquem@physics.ox.ac.uk}, Enrico Martinez$^{2}$\thanks{enrico.martinez@univ.ox.ac.uk} and Alexander Boone$^{3,4}$\thanks{alexander.boone@physics.ox.ac.uk} \\
$^1$ Rudolf Peierls Centre for Theoretical Physics, University of Oxford, 
Parks Road, Oxford OX1 3PU, UK \\
$^2$ University College, High Street, Oxford OX1 4BH, UK\\ 
$^3$ Department of Physics, Astrophysics, University of Oxford, Keble Road, Oxford OX1 3RH, UK\\
$^4$ Somerville College, Woodstock Road, Oxford OX2 6HD, UK  \\
}

\date{Accepted XXX. Received YYY; in original form ZZZ}

\pubyear{2026}

\begin{document}
\label{firstpage}
\pagerange{\pageref{firstpage}--\pageref{lastpage}}
\maketitle

\begin{abstract}
 We present three-dimensional Dedalus simulations of Rayleigh–B\'enard convection with a blackbody-radiating free upper surface, subject to a low-amplitude oscillatory forcing that mimics  tidal perturbations in convective envelopes of stars and planets. The forcing period  is 10–100 times shorter than the convective timescale, $t_{\rm conv}$.
 Using a Reynolds decomposition of the velocity field averaged over one oscillation period, in which the tidal oscillations naturally constitute the fluctuating field and convection the mean flow, we elucidate the kinetic energy exchange between the two.
Provided the oscillatory Reynolds number exceeds a modest threshold, we find that the oscillations systematically transfer kinetic energy to the mean flow at a volume-averaged rate $D_R \sim  u'^2   t_{\rm conv}^{-1}$, where $u'$ is the rms fluctuation velocity.  This reflects strong, order-unity correlations between the fluctuation velocities and the mean flow.  These arise because the oscillatory forcing displaces fluid elements that are then redirected by  buoyancy and incompressibility in the same manner as the mean flow.
The  transfer is dominated by correlations involving vertical velocity fluctuations and vertical  gradients of the  mean flow.  
The resulting energy transfer rate is consistent, within the equilibrium-tide framework, with the observed  tidal circularisation of solar-type binaries and with the orbital evolution of  moons of Jupiter and Saturn.  This validates the formalism proposed by \citet{Terquem2021} for the dissipation of fast tides, a longstanding problem.
Replacing the free surface with a rigid upper boundary significantly and artificially modifies the correlations.  

\end{abstract}



\section{Introduction}

Tidal dissipation in convective flows plays a key role in the evolution of stellar binaries, star-planet systems and the satellite systems of giant planets \citep{Ogilvie2014}. In many of these systems, the tidal forcing varies on timescales much shorter than the convective turnover time. How such fast tides inte\-ract with turbulent convection, and how efficiently they are dissipated, remains an open problem.

Starting with \citet{Zahn1966}, tidal dissipation in convective regions was traditionally described using an effective turbulent viscosity acting on the large-scale oscillatory flow. In this model, based on mixing length theory, dissipation is assumed to arise from
correlations between  convective velocity fluctuations and gradients of the tidal flow.
When the tidal period is much shorter than the convective overturning time, convective eddies are assumed to respond inefficiently to the rapidly varying shear. This behaviour is implemented through a pres\-cribed reduction of the effective turbulent viscosity at short forcing periods (\citealt{Zahn1966}, \citealt{Goldreich1977}), rather than emerging directly from the mixing-length forma\-lism itself.  This leads to dissipation rates that are typically orders of magnitude too small to explain the observations.

\citet{Terquem2021,Terquem2023} showed that this discrepancy reflects a more fundamental limitation of the turbulent-viscosity approach in the fast-tide regime.   When the forcing period is short, the tidal flow must instead be treated as part of the fluctuating velocity field, with the mean flow defined by temporal averaging over an oscillation period.  Dissipation then arises from correlations between the oscillatory velocity fluctuations and gradients of the convective flow, rather than from convective motions being sheared by the tide.  This leads to a qualitatively different dissipation law: the rate still scales with the inverse convective timescale but, crucially, it does not suffer the strong reduction at short forcing periods that is artificially imposed in traditional turbulent-viscosity mo\-dels.

Despite this theoretical progress, the existence and robustness of such Reynolds-stress correlations in fully three-dimensional convection remain debated. 
Most numerical studies have instead focused on estimating an effective turbulent viscosity  \citep{Penev2009, Ogilvie2012, Duguid2020, Vidal2020a, Vidal2020b}.  
To date, the only direct numerical attempt to evaluate the dissipation rate proposed by \citet{Terquem2021}  in a fully convective flow is that of \citet{Barker2021}.  However, the use of rigid boundaries at both the top and bottom of the domain precludes a fully  consistent computation of  the work done by a potential tidal force in an incompressible fluid. As a consequence, they prescribe the tidal response  kinematically as an irrotational velocity field, appearing only through advection terms
 in the mean-flow equations. While this setup captures some aspects of tide-convection interactions, it does not allow the oscillatory tidal flow to emerge self-consistently as the dynamical response to an external forcing. 

In contrast, the present study models tides as an externally applied oscillatory forcing, allowing the full velocity field to adjust dynamically and to generate secondary fluctuations as part of the convective flow itself. This distinction is crucial: the Reynolds-stress correlations responsible for energy transfer arise precisely because oscillatory displacements are redirected by buoyancy and incompressibility in the same way as the mean flow itself, rather than being prescribed \textit{a priori} as a fixed velocity field. The forcing period is chosen to be one to two orders of magnitude shorter than the convective timescale, placing the system firmly in the fast-tide regime. 
In this limit, the mean flow is naturally  defined by averaging over the oscillation period rather than over the convective timescale. 
Accordingly, we adopt a Reynolds decomposition based on temporal averaging over one oscillation period and analyse the kinetic-energy exchange between the oscillatory fluctuations and the mean convective flow.

To the best of our knowledge, this is the first three-dimensional numerical study of turbulent Rayleigh-B\'enard convection with a dynamically deformable free surface and radiative cooling, subject to fast oscillatory forcing. 
Previous numerical studies of convection with deformable free surfaces have largely focused on weakly nonlinear Marangoni-driven or mixed Rayleigh-B\'enard-Marangoni regimes, primarily concerned with instability and pattern formation, rather than on turbulent convection and the energy transfer between oscillatory motions and the mean flow considered here (e.g., \citealt{Cliffe1998}, \citealt{Dang2021}).
 The free surface is essential for capturing the large-scale vertical structure of convective plumes and the associated mean-flow gradients that drive the Reynolds-stress correlations.   As we show below, imposing a rigid upper boundary qualitatively alters these correlations and can suppress the net energy transfer altogether. Moreover, when the flow is incompressible and the forcing derives from a scalar potential, a  free surface is required for the forcing  to perform net work on the flow.

We show that, above a modest threshold in the fluctua\-ting Reynolds number, the oscillations systematically transfer kinetic energy to the mean flow at a rate
$D_R \sim  u'^2 \, t_{\rm conv}^{-1}$, 
independent of the forcing period over the range explored, where $u'$ is the rms fluctuation velocity.  We identify the Reynolds-stress components responsible for this transfer and discuss how viscous diffusion, thermal diffusion and boundary conditions control the establishment of the underlying correlations. Our results provide direct numerical support for the fast-tide dissipation mechanism proposed by \citet{Terquem2021,Terquem2023}.

The outline of the paper is as follows.
In Section~\ref{sec:equations}, we introduce the governing equations, define the equilibrium state and describe the Reynolds decomposition appropriate for fast oscillatory forcing.  We also derive the mean-flow and fluctuation momentum equations, together with the associated local and global kinetic-energy balances that form the theoretical framework of the study.  Section~\ref{sec:BC} presents the boundary conditions, including the formulation of the radiating, deformable free surface adopted here. In Section~\ref{sec:simulations}, we describe the numerical setup. We first present unforced reference simu\-lations used to characterise the convective flow, and then introduce the oscillatory forcing. We consider both a purely vertical forcing, which provides a simple framework for understanding how correlations arise within the convective flow, and a forcing that derives from a scalar potential and mimics tidal forcing.  We then analyse the resulting energy budgets and Reynolds-stress transfer, and examine their dependence on viscosity and boundary conditions.   In Section~\ref{sec:discussion}, we discuss the physical origin of the Reynolds-stress correlations.  Finally, Section~\ref{sec:summary} summarises the main results and discusses their implications for tidal dissipation in stellar and planetary convective envelopes.

\section{Governing equations} 
\label{sec:equations}

We consider a fluid with velocity ${\bf u}$,  mass density $\rho$, pressure $P$ and temperature $T$ subject to an external gravitational field $\bf g$ and an imposed force per unit mass ${\bf f}$.  The fluid has constant 
kinematic viscosity $\nu$ and thermal diffusivity $\kappa$.  We adopt Cartesian coordinates $\left( x, y, z \right)$ with corresponding unit vectors $\left( \hat{\bf x} ,\hat{\bf y} ,\hat{\bf z} \right)$,
such that ${\bf g} = -g \hat{\bf z}$.    The force ${\bf f}$ represents an externally imposed perturbation and is taken to be zero in the equilibrium state defined below.

The system is governed by the continuity, Navier-Stokes and thermal 
energy equations:

\begin{equation}
\frac{\partial \rho}{\partial t} +   \nab \cdot \left( {\rho \bf u} \right)= 0,
\label{eq:mass}
\end{equation}

\begin{equation}
 \frac{\partial {\bf u}}{\partial t} +  {\bf u} \cdot \nab {\bf u} = - \frac{1}{\rho} \nab P +  \nu \nabla^2 {\bf u} - g \hat{\bf z} + {\bf f} ,
\label{eq:NS}
\end{equation} 

\begin{equation}
 \frac{\partial T}{\partial t} +  {\bf u}\cdot \nab T = \kappa \nabla^2 T .
 \label{eq:thermal}
\end{equation}

\subsection{Equilibrium state}

In equilibrium, the fluid is at rest (${\bf u}={\bf 0}$), the external forcing vanishes (${\bf f}={\bf 0}$) and the domain spans $0 \le z \le H$. The tempe\-rature is fixed at $T_1$ at the bottom and $T_2$ at the surface, with $T_2<T_1$.  We define  $\Delta T \equiv T_1-T_2$ and assume that the interior temperature profile is linear:

\begin{equation}
T_0 \left(z \right) = T_1 - \frac{  T_1 - T_2 }{H} z.  
 \label{eq:thermaleq}
\end{equation}

\noindent With this choice, the thermal equation~(\ref{eq:thermal}) is automatically satisfied.  The mass density is taken to be uniform, equal to $\rho_0$, and the equilibrium pressure profile $P_0 \left(z \right)$ satisfies the hydrostatic balance condition from equation~(\ref{eq:NS}):

  \begin{equation}
  \nab P_0 =  - \rho_0 g \hat{\bf z} .
\label{eq:NSeq}
\end{equation}

\subsection{Perturbation and Boussinesq approximation}

Deviations from the equilibrium state are characterized by a velocity field ${\bf u}= \left( u_x, u_y, u_z \right)$ and perturbations in density, pressure and temperature: $\rho = \rho_0 + \delta \rho$,  $P = P_0 + \delta P$ and $T = T_0 + \delta T$.   We assume that density variations are small and arise solely from temperature
perturbations, with pressure-induced density variations neglected. Specifically, we write:

\begin{equation}
\rho= \rho_0 \left(  1 - \alpha \delta T    \right),
\label{eq:deltarho}
\end{equation}   

\noindent where $\alpha$ is the (constant) coefficient of thermal expansion and 
$\left| \alpha \delta T \right| \ll 1$.   Within the Boussinesq approximation, density  variations are neglected in the mass conservation equation~(\ref{eq:mass}), yielding the incompressibility condition:

\begin{equation}
 \nab \cdot {\bf u} = 0.
\label{eq:massp} 
\end{equation}

\noindent In the momentum equation~(\ref{eq:NS}), density variations affect the dynamics only through the pressure-gradient term.   A first-order expansion of $1/\rho$  in $\delta T$, combined with the hydrostatic balance~(\ref{eq:NSeq}),   produces the buoyancy term $g \alpha\delta T \hat{\bf z}$.
The go\-ver\-ning equations for the perturbed quantities then become:

\begin{equation}
 \frac{\partial {\bf u}}{\partial t} + {\bf u} \cdot \nab {\bf u} = - \frac{1}{\rho_0} \nab \delta P +  \nu \nabla^2 {\bf u} +  g \alpha \delta T \hat{\bf z} +  {\bf f} ,
\label{eq:NSp} 
\end{equation}

\begin{equation}
  \frac{\partial \delta T}{\partial t} + {\bf u} \cdot \nab \left( T_0 + \delta T \right)  = \kappa \nabla^2 \delta T .
 \label{eq:thermalp}
\end{equation} 

\noindent In the absence of external forcing (${\bf f}={\bf 0}$), this system reduces to the classical Rayleigh-B\'enard problem.  Here, we focus on the turbulent convective regime (unstable equilibrium) and apply ${\bf f}$  as an oscillatory forcing to model  perturbations within an already convective flow. 

\subsection{Dimensionless form}
\label{sec:dimensionlessform}

The Rayleigh and Prandtl numbers are defined as:

\begin{equation}
Ra = \frac{\alpha \Delta T g d^3}{\nu \kappa}, \qquad Pr = \frac{\nu}{\kappa},
\end{equation}

\noindent where  $d$ is a characteristic length scale.  
The Rayleigh number measures the relative strength of buoyancy-driven advection compared to thermal and viscous  diffusion.
When $Ra$ exceeds a critical  value $Ra_c$, 
 the equilibrium state becomes unstable and thermal convection develops, corresponding to the  classical Rayleigh-B\'enard instability.   In non-turbulent regimes, $d$ is typically taken to be the depth of the convective layer (here, $H$). However,  in turbulent flows,  a more appropriate length scale may be the pressure scale height.  We  therefore formulate the equations using  a general characteristic length $d$. For the numerical simulations presented below, where the pressure scale height and layer depth are comparable, we take $d = H$.
 
 \noindent We also define dimensionless viscosity and thermal diffusivity  as:
 
 \begin{align}
 \tilde{\kappa} &  \equiv  \frac{\kappa}{\sqrt{\alpha \Delta T g d^3}} =  Ra^{-1/2} Pr^{-1/2} ,  
 \label{eq:kappa}
 \\
 \tilde{\nu} & \equiv   \frac{\nu}{\sqrt{\alpha \Delta T g d^3}} =   Ra^{-1/2} Pr^{1/2} .
 \label{eq:nu}
 \end{align}

\noindent Next, we introduce a characteristic velocity and timescale:

\begin{equation}
U=\sqrt{\alpha \Delta T g d},  \qquad  \tau = \frac{d}{U} = \sqrt{\frac{d}{ \alpha \Delta T g} }. 
\end{equation}

\noindent Using these, we define the following dimensionless variables:
velocity $\tilde{\bf u} = {\bf u}/U$, time $\tilde{t}=t/\tau$, pressure $\tilde{P}=\delta P / \left(  \rho_0 U^2 \right)$, temperature perturbation $\tilde{b}=\delta T / \Delta T $ and external force $\tilde{\bf f} = {\bf f} / \left( \alpha \Delta T g \right)$.   
We also rescale the spatial coordinates as $\tilde{x}=x/d$, $\tilde{y}=y/d$, $\tilde{z}=z/d$, which gives  the dimensionless gradient  operator $\tilde{\nab} =  d \nab$.

Substituting these expressions into equations~(\ref{eq:massp}), (\ref{eq:NSp}) and~(\ref{eq:thermalp}) yields the  dimensionless equations:

\begin{equation}
 \tilde{\nab} \cdot \tilde{\bf u} = 0 ,
\label{eq:massdim} 
\end{equation}

\begin{equation}
 \frac{\partial \tilde{\bf u}}{\partial \tilde{t}} + \tilde{\bf u} \cdot \tilde{\nab} \tilde{\bf u} = -\tilde{\nab} \tilde{P} +  \tilde{\nu} \tilde{\nabla}^2 \tilde{\bf u} +  \tilde{b} \hat{\bf z}  +   \tilde{\bf f},
\label{eq:NSdim} 
\end{equation} 

\begin{equation}
  \frac{\partial \tilde{b}}{\partial \tilde{t}} + \tilde{\bf u} \cdot \tilde{\nab} \tilde{b} - \tilde{u}_z   = \tilde{\kappa} \tilde{\nabla}^2 \tilde{b} .
 \label{eq:thermaldim}
\end{equation}

\noindent The quantity $b \equiv \delta T$ is often loosely referred to as the {\em buo\-yan\-cy}, although the actual buoyancy force per unit mass is $g \alpha \delta T$.  

We also introduce the dimensionless stress tensor:

\begin{equation}
\tilde{\sigma}_{ij}= \tilde{\sigma}_{ji}= \tilde{\nu} \left( \frac{\partial \tilde{u}_i}{\partial \tilde{x}_j} +   \frac{\partial \tilde{u}_j}{\partial \tilde{x}_i} \right)  ,
\label{eq:stressdim}
\end{equation}

\noindent where indices $i $ and $j$ take values in $(1,2,3)$, corresponding to $x_1=x$, $x_2=y$ and $x_3=z$.
Because the flow is incompressible,   the viscous term in 
equation~(\ref{eq:NSdim}) may equivalently be written as $ \tilde{\nu} \tilde{\nabla}^2 \tilde{ u}_i  = \partial \tilde{\sigma}_{ij} / \partial \tilde{x}_j $.

\subsection{Timescales and Reynolds decomposition}
\label{sec:timescales}

In stellar models, the convective timescale $t_{\rm conv}$ at a radius $r$ is the time required for energy to be transported across the local mixing length (eddy turnover length).  It is commonly defined as  $t_{\rm conv} \sim H_P/V_{\rm conv}$, where $H_P$ is the local pressure scale height and $V_{\rm conv}$ is the horizontally averaged convective velocity on the sphere of radius $r$.  The Rayleigh-B\'enard simulations presented here model  a local patch of a stellar convective envelope, with the vertical extent of the domain corresponding to the eddy turnover length.  Accordingly, we define the convective timescale in the volume ${\mathcal V}$ of the simulation as: 
\begin{equation}
t_{\rm conv} = \frac{H}{ u_{z, \rm rms}} ,
\label{eq:tconv}
\end{equation}
where: 
\begin{equation}
u_{z, \rm rms} = \sqrt{ \frac{1}{\mathcal V} \int_ {\mathcal V} u_z^2 \; {\rm d} v } ,
\label{eq:uzrms}
\end{equation}
is the volume-averaged rms  vertical velocity.    Both upward and downward motions contribute to $u_{z,{\rm rms}}$, consistent with the fact that energy is transported towards the surface by the correlation between $ u_z$ and the temperature fluctuation $\delta T$, not by the mean vertical flow.

Another relevant timescale is the crossing time:
\begin{equation}
 t_{\rm cross} = \frac{H}{u_{z, {\rm max}} } , 
 \label{eq:tcross}
 \end{equation}
 defined as  the time it takes for the fastest plumes to traverse the layer.  A defining feature of convection, especially at high Rayleigh numbers,  is that $t_{\rm cross} \ll t_{\rm conv}$, because heat transport is dominated by rare, rapid plumes that occupy only a small fraction of the flow volume.

We consider a force ${\bf f}$ that oscillates in time with  period $t_{\rm osc}$  such that $t_{\rm osc} \ll t_{\rm conv}$.   We perform a Reynolds decomposition: 

\begin{equation}
{\bf u} = {\bf V} + {\bf u}', \quad {\rm where} \quad {\bf V} \equiv \left< {\bf u} \right>
\label{eq:Reynolds}
\end{equation}
with the angle brackets denoting a time average over one oscil\-lation period $t_{\rm osc}$.  By construction, at a  fixed point in the  flow, $ {\bf u}'$ oscillates about zero on the timescale $t_{\rm osc}$.   The mean velocity ${\bf V}$ itself evolves on a much longer timescale $\tau_V$  determined by the intrinsic convective dynamics.  At high Rayleigh numbers, where convection is fully turbulent,  $\tau_V \sim t_{\rm cross}$.  In contrast, at the moderate to low Rayleigh numbers explored in this study, convective structures are more persistent and  $\tau_V > t_{\rm conv}$, so that the mean flow appears quasi-steady on timescales  small compared to $t_{\rm conv}$.

In three dimensions, convection supports a broad spectrum of eddies with varying velocities, lengthscales and  timescales, reflecting the cascade of kinetic energy from large to small scales.  
Since the largest eddies  have both the highest velocities and the longest intrinsic timescales, the mean flow ${\bf V}$ is dominated by these large-scale coherent structures. The fluctuation field ${\bf u}'$ comprises (i) the direct velocity response to the oscillating force, (ii) secondary oscillations generated by the interaction of this response with convection,  and (iii) convective motions whose intrinsic timescales are comparable to or shorter than $t_{\rm osc}$.

\subsection{Averaged equations}

We  focus on cases where $t_{\rm osc} \ll t_{\rm conv}$.
In this regime, mean quantities can be considered approximately constant over a single oscillation period.  This assumption remains valid even when  $t_{\rm osc}$ is comparable to, or exceeds,  $t_{\rm cross}$, provided that $t_{\rm cross} \ll t_{\rm conv}$.  

Substituting the Reynolds decomposition into the incompressibility equation~(\ref{eq:massdim}) and averaging over $\tilde{t}_{\rm osc}$ gives:

\begin{equation}
 \tilde{\nab} \cdot \tilde{\bf V} = 0.
\label{eq:massdim_mean} 
\end{equation}

\noindent Subtracting this from equation~(\ref{eq:massdim})  yields:

\begin{equation}
 \tilde{\nab} \cdot \tilde{\bf u}' = 0,
\label{eq:massdim_fluc} 
\end{equation}

\noindent indicating that both the mean flow and the fluctuations are incompressible.  

\subsubsection{Momentum equations}
\label{sec:momentum}

Substituting the Reynolds decomposition into the momentum equation~(\ref{eq:NSdim}) and averaging over $\tilde{t}_{\rm osc}$ gives the mean-flow momentum equation:

\begin{align}
 \frac{\partial \tilde{ V}_i}{\partial \tilde{t}} + \tilde{V}_j \frac{ \partial  \tilde{ V}_i}{\partial \tilde{x}_j} +  &  \left<  \tilde{ u}'_j  \frac{ \partial   \tilde{ u}'_i }{\partial \tilde{x}_j} \right> =
 \nonumber \\
&  -\frac{\partial \left<  \tilde{P}\right> }{\partial \tilde{x}_i}  +  \frac{\partial \left< \tilde{  \sigma}_{ij} \right> }{\partial \tilde{x}_j} +  \left< \tilde{b} \right> \delta_{3i}   , 
 \label{eq:NSmean}
\end{align} 

\noindent with summation over repeated indices implied.  We have used the identity $\left< \partial \tilde{ u}'_i / \partial \tilde{t} \right> = 0$, which holds because $\tilde{u}'_i$ is periodic over the averaging time interval.

Equation~(\ref{eq:NSmean}) retains all nonlinear terms and does not assume that the fluctuations  are small compared to the mean quantities.  Since the external forcing $\tilde{\bf f}$  does not appear explicitly, it influences the mean flow only through the term  $\left<  \tilde{u}'_j  \left( \partial  \tilde{u}'_i    / \partial \tilde{x}_j  \right) \right>$.  
As long as this term remains small compared to the dominant buoyancy contribution $\left< \tilde{b} \right>$
 and/or the mean advection term  $ \tilde{V}_j  \left(  \partial  \tilde{ V}_i / \partial \tilde{x}_j \right) $,
 the amplitude of the forcing may be chosen freely without altering the
leading-order balance of the mean flow.


\noindent Subtracting equation~(\ref{eq:NSmean}) from equation~(\ref{eq:NSdim}) gives the fluctuation momentum equation:

\begin{align}
\frac{\partial \tilde{ u}'_i}{\partial \tilde{t}} +   \tilde{u}'_j \frac{ \partial  \tilde{ V}_i}{\partial \tilde{x}_j} +   \tilde{V}_j \frac{ \partial  \tilde{ u}'_i}{\partial \tilde{x}_j} +   \tilde{ u}'_j \frac{ \partial \tilde{ u}'_i}{\partial \tilde{x}_j}   &  -   \left<  \tilde{ u}'_j \frac{ \partial \tilde{ u}'_i}{\partial \tilde{x}_j} \right>  = 
 \nonumber \\
&  -\frac{\partial   \tilde{P}' }{\partial \tilde{x}_i}  +  \frac{\partial  \tilde{  \sigma}'_{ij}  }{\partial \tilde{x}_j} +   \tilde{b}'  \delta_{3i} +  \tilde{ f}_i  , 
 \label{eq:NSfluc}
\end{align} 

\noindent where the fluctuating components are defined as: $\tilde{P}'= \tilde{P} - \left< \tilde{P} \right>$, $\tilde{b}'= \tilde{b} - \left< \tilde{b} \right>$ and 
$\tilde{\sigma}'_{ij}= \tilde{\sigma}_{ij} - \left< \tilde{\sigma}_{ij} \right>$.

\subsubsection{Local energy equations}

Multiplying equation~(\ref{eq:NSmean}) by $\tilde{V}_i$ and summing  over $i$ yields the  kinetic energy equation for the mean flow:

\begin{equation}
  \dfrac{\partial \tilde{E}_k}{\partial \tilde{t}}  = 
   -   \dfrac{\partial \tilde{F}_j}{\partial \tilde{x}_j}    
    + \tilde{D}_R - \tilde{D}_v +  \left< \tilde{b} \right> \tilde{V}_z   ,
 \label{eq:KEmean}
\end{equation} 

\noindent where $\tilde{E}_k \equiv \tfrac{1}{2} \left( \tilde{V}_x^2+\tilde{V}_y^2+\tilde{V}_z^2 \right) $ is the kinetic energy of the mean flow. The transfer and dissipation terms
are given by:

\begin{align}
\tilde{D}_R & =   \left< \tilde{ u}'_i  \tilde{ u}'_j  \right> \frac{ \partial \tilde{V}_i }{\partial \tilde{x}_j} , \\
\tilde{D}_v & =  \left<  \tilde{  \sigma}_{ij}  \right> \frac{ \partial \tilde{V}_i }{\partial \tilde{x}_j} .
\end{align}

\noindent The associated mean energy flux $\tilde{\bf F}$ is:

\begin{equation}
\tilde{F}_j  = \frac{ \tilde{V}^2 \tilde{V}_j }{2}   
 +  \left< \tilde{ u}'_i  \tilde{ u}'_j  \right> \tilde{V}_i -  \left< \tilde{  \sigma}_{ij} \right> \tilde{V}_i +  \left< \tilde{P} \right> \tilde{V}_j .
 \end{equation}

\noindent In deriving equation~(\ref{eq:KEmean}), we have used the incompressibility of both $\tilde{\bf V}$ and $\tilde{\bf u}'$.  For later reference, we define the kinetic contribution to the flux as:
\begin{equation}
\tilde{F}_{{\rm kin}, j}  = \frac{ \tilde{V}^2 \tilde{V}_j }{2}   
 +  \left< \tilde{ u}'_i  \tilde{ u}'_j  \right> \tilde{V}_i .
\label{eq:kineticflux} 
 \end{equation}

\noindent Similarly, multiplying equation~(\ref{eq:NSfluc}) by $\tilde{u}'_i$, summing over $i$ and averaging over $\tilde{t}_{\rm osc}$ yields the kinetic energy equation for the fluctuations:

\begin{equation}
  \dfrac{\partial \left< \tilde{e}'_k \right>}{\partial \tilde{t}}   = 
   -   \frac{ \partial \left< \tilde{F}'_j \right> }{\partial \tilde{x}_j} 
    - \tilde{D}_R - \tilde{D}'_v +  \left< \tilde{b}' \tilde{u}'_z \right>  
       + 
    \left< \tilde{\bf f} \cdot \tilde{\bf u}' \right>   ,
  \label{eq:KEfluc}
\end{equation} 

 \noindent where $\tilde{e}'_k  \equiv \tfrac{1}{2} \left(  \tilde{u}'^2_x + \tilde{u}'^2_y +\tilde{u}'^2_z \right) $ is the kinetic energy of the fluctuations. The viscous dissipation rate is:

\begin{equation}
\tilde{D}'_v  =  \left<  \tilde{  \sigma}'_{ij}   \frac{ \partial \tilde{u}'_i }{\partial \tilde{x}_j} \right> ,
\label{eq:Dvfluc}
\end{equation}

\noindent and the mean fluctuation energy flux satisfies:

\begin{equation}
\left<  \tilde{F}'_j \right>= \frac{ \left< \tilde{u}'^2 \right> \tilde{V}_j }{2}   
 +  \frac{\left< \tilde{u}'^2 \tilde{u}'_j \right> }{2}  -   \left<  \tilde{  \sigma}'_{ij} \tilde{u}'_i  \right> +  \left< \tilde{P}' \tilde{u}'_j \right> .
 \label{eq:flucflux}
\end{equation}

\noindent In deriving equation~(\ref{eq:KEfluc}), we have used the identity $\left< \partial \tilde{ e}'_k / \partial \tilde{t} \right> =  \partial \left< \tilde{ e}'_k \right> / \partial \tilde{t} $, which holds because the averaging is performed over one oscillation cycle of
$\tilde e'_k$.

The term $\tilde{D}_R$ represents the rate of energy exchange between the mean flow and the fluctuations via the Reynolds stress, i.e. correlations between the components of the fluctuation velocity.   When $\tilde{D}_R>0$, energy is transferred from the fluctuations to the mean flow.  The objective  of this study is to determine the sign of $\tilde{D}_R$.
We emphasise that $\tilde{D}_R$ is a {\em transfer term}, not a dissipation rate. 
When $\tilde{D}_R>0$, the fluctuations are effectively damped because they lose energy to the mean flow.  However, this energy is not dissipated directly at the fluctuation scale.  Instead, it is subsequently removed either by viscosity acting on the mean flow or by transport through the boundaries {\em via} the energy flux. 

\subsubsection{Global energy balances}

Integrating equation~(\ref{eq:KEmean}) over the flow volume yields the global energy balance for the mean flow:

\begin{equation}
- \int \tilde{\bf F} \cdot \hat{\bf n} \; {\rm d} s + \int  \Big( \tilde{D}_R - \tilde{D}_v +  \left< \tilde{b} \right> \tilde{V}_z  \Big) {\rm d} v  =0  ,
\label{eq:globalenergymean}
\end{equation}

\noindent where the surface integral  is taken over the boundaries of the domain, with  $\hat{\bf n}$ the outward unit normal.  We have assumed a statistically steady state ($ \partial  \tilde{E}_k / \partial \tilde{t}  =0 $), although the time variation can be retained if necessary. 

Similarly, integrating equation~(\ref{eq:KEfluc}) gives the global energy balance for the fluctuations:

\begin{equation}
 - \int \left< \tilde{\bf F}' \right>  \cdot \hat{\bf n} \; {\rm d} s 
 + \int \Big( - \tilde{D}_R - \tilde{D}'_v + \left< \tilde{b}' \tilde{u}'_z \right> 
 + \left< \tilde{\bf f} \cdot \tilde{\bf u}' \right>
 \Big)  {\rm d} v = 0 .
\label{eq:globalenergyfluc}
\end{equation}

\noindent Again,  the time variation of $\left< \tilde{e}'_k \right>$ may be included if needed.   

\section{Boundary conditions}
\label{sec:BC}

Periodic boundary conditions are applied in the horizontal directions.

\subsection{Lower boundary ($z=0$)}

By analogy with the Sun, the convectively unstable region considered here may overlie a stably stratified layer, with $z=0$ representing the interface between the two.  In such configurations, convective overshooting can lead to a nonzero vertical velocity at this boundary.  However, this effect is not included in the present study, and we therefore impose $\tilde{u}_z=0$ at $z=0$.  

To complete the velocity boundary conditions,  we assume that the tangential stresses vanish at the interface: 

\begin{equation}
\tilde{\sigma}_{xz} = 0, \quad  \tilde{\sigma}_{yz} = 0.
\end{equation}

\noindent    Such a stress-free interface approximates the behaviour of a fluid-fluid boundary with negligible resistance to shear.  

For the temperature, we impose a fixed-flux condition at the lower boundary, corresponding to constant heating from below.  The radiative flux at $z=0$ is held equal to its hydrostatic equilibrium value, denoted $F_0$:

\begin{equation}
-k \nab T \cdot \hat{\bf n} = F_0 ,
\end{equation}

\noindent where $k$ is the thermal conductivity and $\hat{\bf n}$ is the unit normal to the boundary.  At $z=0$, we have $\hat{\bf n} = \hat{\bf z}$, so the condition becomes $\partial T / \partial z = - F_0 /k$.  Writing $T=T_0 + \delta T$, and noting that  $F_0$ sets the equilibrium gradient ${\rm d} T_0 / {\rm d} z = - F_0 / k$, we obtain the final boundary condition:

\begin{equation}
\frac{\partial \tilde{b}}{\partial \tilde{z}} = 0 . 
\end{equation}

\subsection{Free surface ($z=H+\eta$)}
\label{sec:freesurface}

In  the simulations presented below, the upper surface is free and located at $z= H + \eta \left( x, y \right) $, where $\eta$ represents  the departure from the equilibrium height $z=H$.  The surface radiates as a blackbody at  temperature $T \left( H+ \eta \right)$.  

In appendix~\ref{appendixA}, we derive  the kinematic condition~(\ref{eq:etasurf}), the continuity of tangential stresses~(\ref{eq:stressx}) and~(\ref{eq:stressy}), the continuity of  normal stress~(\ref{eq:stressnorm})   and the radiative condition~(\ref{eq:surfrad}). Together, these form the surface boundary conditions,  valid to first order in $\left| \eta \right| / H$.  

These equations include terms involving products of $\tilde{\eta}=\eta/d$ with dimensionless perturbed quantities.  Since $\left| \tilde{\eta} \right| \ll 1$ in all simulations presented here,  such nonlinear terms are neglected in comparison to the  perturbations themselves.  This leads to the following boundary conditions at $\tilde{z}=L_z \equiv H/d$:

\begin{equation}
\tilde{u}_z= \frac{\partial \tilde{\eta}}{\partial \tilde{t}} ,
\end{equation}

\begin{equation}
\tilde{\sigma}_{xz}= 0 ,  \quad \tilde{\sigma}_{yz}= 0 ,
\end{equation}

\begin{equation}
 \tilde{\sigma}_{zz} + \frac{\tilde{\eta}}{\alpha \Delta T } - \tilde{P} = 0,
 \label{eq:sigmazz}
\end{equation}

\begin{equation}
   \frac{\partial \tilde{b}}{\partial \tilde{z}} = 1 + 4 \frac{\tilde{\eta}}{\tilde{T}_2}  -  \left( 1 +   \frac{\tilde{b}}{\tilde{T}_2}\right)^4 .
\end{equation}

\section{Numerical simulations}
\label{sec:simulations}

This section describes the numerical setup and diagnostics used to quantify the transfer of kinetic energy between fluctuations and the mean flow. We first discuss the nondimensional parameters, present the  time-averaging procedure used to separate the mean flow from the fluctuations and characterise the unforced convective state that serves as the baseline reference.  We then introduce an external oscillatory forcing and discuss the conditions required to ensure it probes the coupling between fluctuations and the mean flow  without altering the leading-order dynamics of the convection.  Finally, we analyse the forced simulations through their global energy budgets, dissipation scales and the detailed structure of the Reynolds-stress transfer term $D_R$, demonstrating its robustness  across different oscillation timescales and forcing configurations.

We solve equations~(\ref{eq:massdim}),  (\ref{eq:NSdim}) and~(\ref{eq:thermaldim}) using the spectral code Dedalus \citep{Burns2020}.  For our Cartesian Rayleigh–B\'enard setup, the horizontal directions are expanded in Fourier bases and the vertical direction in Chebyshev polynomials.  

\subsection{Parameter choices and nondimensional numbers}

As discussed in section~\ref{sec:dimensionlessform},  we adopt $d=H$ (the vertical extent of the domain) as the characteristic length scale, yiel\-ding $L_z \equiv H/d = 1$. The horizontal dimensions are  $L_x=L_y=3$, sufficient to accommodate two convection 
cells while 
remai\-ning computationally tractable. We set the Prandtl number  to $Pr=1$ and, in most  simulations, the Rayleigh number to  $Ra=10^6$, corresponding to $\tilde{\nu} = \tilde{\kappa} = 10^{-3}$.  These are the only dimensionless control parameters entering the governing equations.   For this parameter set and the range of oscillation periods $t_{\rm osc}$ considered, numerical convergence of the forced simulations required 160–192 grid points in the vertical direction and 60–70 grid points in each horizontal direction.

For the boundary conditions, two additional quantities must be specified: $\alpha \Delta T$ and  $\tilde{T}_2$.
The product $\alpha \Delta T$ represents the fractional density change across the imposed temperature difference $\Delta T$ and must be small to justify the 
Boussinesq  approximation. We therefore set $\alpha \Delta T = 0.1 $.  

Because the flow is heated from below, convection cools the lower region and warms the upper boundary layer through the upward transport of heat.  As a result,  the horizontally averaged temperature perturbation $\delta T$  at the upper surface is positive, implying a radiative  flux  $\sigma T ^4$ that exceeds the equilibrium value $F_0$.  This boundary condition, which provides a more realistic representation of a stellar envelope or giant-planet atmosphere, makes convection slightly more vigorous than in the case of a fixed-flux upper boundary.  We have verified that this choice does not alter our main conclusions by performing additional simulations with a fixed-flux condition at the top.
 In the simulations presented here, we  fix the surface temperature $\tilde{T}_2=0.25$, corresponding to $T_2/T_1=0.2$.  Varying $\tilde{T}_2$ within a reasonable range does not 
affect the quali\-tative behaviour of the flow.

\subsection{Averaging procedure}
\label{sec:averaging}

To track the energy budget of the fluctuations, we compute time averages over one oscillation period $t_{\rm osc}$ not only of the full fields (e.g. velocity components), but also of  products of fluctuating quantities (e.g. $u'_i u'_j$).  Using equation~(\ref{eq:Reynolds}), we compute the average $Q$ of any quantity $q$ as:

\begin{equation}
Q \left( nt_{\rm osc} \right)  \equiv \left< q \right> \left( nt_{\rm osc} \right)  =\frac{1}{t_{\rm osc}} \int_{n t_{\rm osc}}^{(n+1) t_{\rm osc}} q \, {\rm d}t .
\end{equation}

Consider two quantities $q$ and $r$ whose averages $Q$ and $R$ vary on timescales much longer than $t_{\rm osc}$.  We then have:

\begin{equation}
\left< qr \right> = \left< \left( Q+q' \right) \left( R+r' \right) \right> = QR + \left< q' r' \right>   ,
\end{equation}

\noindent which implies $\left< q' r' \right> = \left< qr \right> - QR.$  This identity allows us to compute $\left< q' r' \right>$ without storing the fluctuating fields $q'$ and $r'$, relying only on time-averaged quantities.  The same decomposition readily extends to products of three or more  fluctuating fields, as needed for the  flux terms in the  energy budgets.  

An important consequence of this finite-window block avera\-ge is that, even in the absence of external forcing, the residual field $u'$ contains contributions from the temporal evolution of the convective flow on timescales longer than $t_{\rm osc}$ within each averaging window. This is distinct from what we would obtain with, say, a Fourier decomposition at the forcing frequency, which would isolate only the oscillatory response. As we will see below, in the forced case, the externally driven oscillations dominate the fluctuation budget and this distinction is inconsequential.  However,  in the unforced case, the fluctuation field is dominated by this finite-window artefact. Keeping this in mind is essential for correctly interpreting the sign and magnitude of $D_R$ in the two cases, as discussed below.

\subsection{Simulations  without forcing}

We begin by considering unforced Rayleigh-B\'enard convection, i.e. by setting $\tilde{\bf f}=0$ in equation~(\ref{eq:NSdim}). 

\begin{figure*}
\centering
\includegraphics[width=1.4\columnwidth,angle=0]{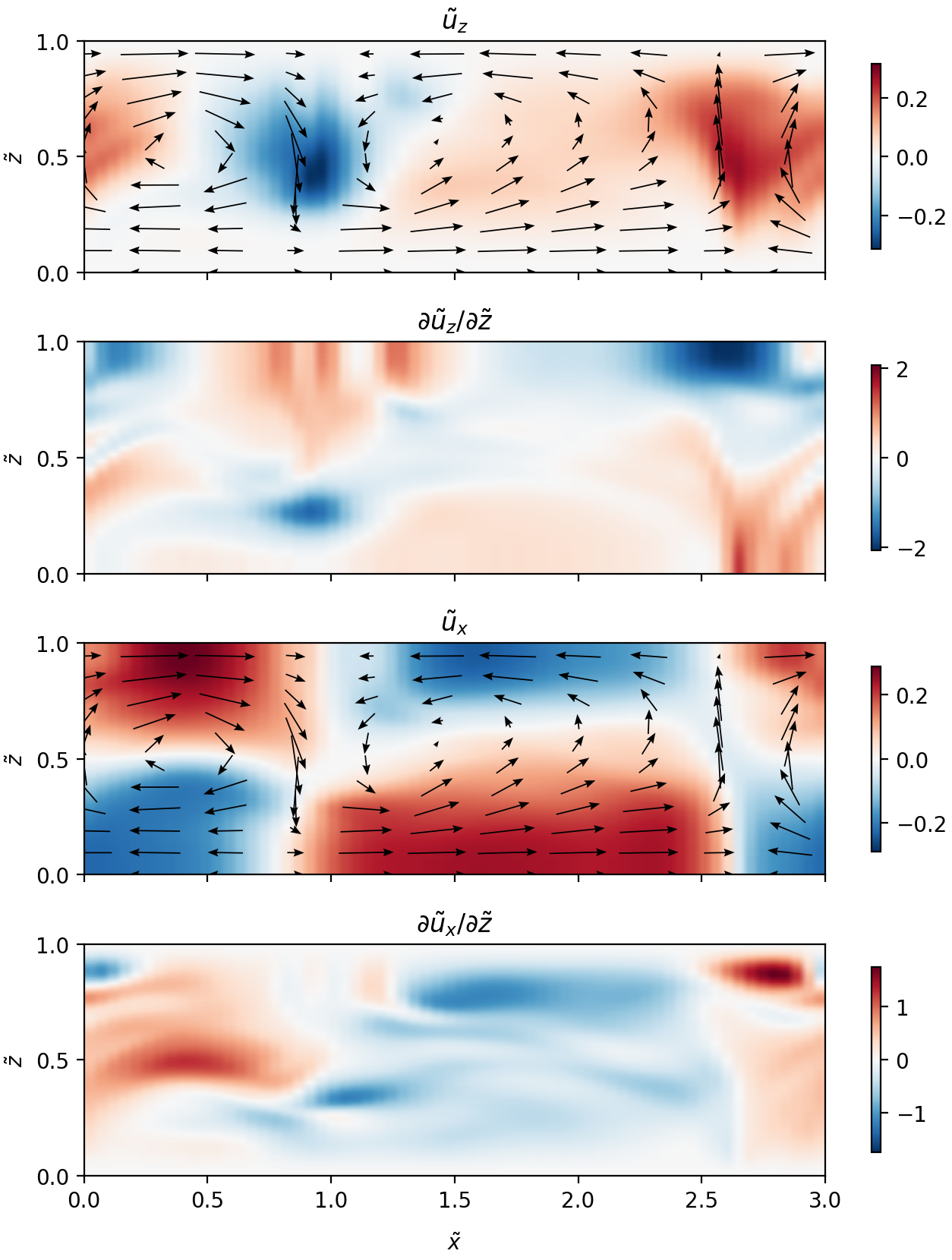}
\caption{Instantaneous snapshots of the velocity field in the plane $\tilde{y}=L_y/2$ at $\tilde{t} \simeq 120$.  From top to bottom,
the panels show $\tilde{u}_z$, $\partial \tilde{u}_z/\partial \tilde{z}$, $\tilde{u}_x$ and $\partial \tilde{u}_x/\partial \tilde{z}$. Colour shading  indicates scalar magnitude (red and blue corresponding to positive and negative values, respectively).  Arrows in the velocity panels represent the in-plane velocity vector  $(\tilde{u}_x,\tilde{u}_z)$, with lengths proportional to the velocity magnitude. Coordinates are shown in units of the domain dimensions, $0 \le \tilde{x} \le L_x$ and $0 \le \tilde{z} \le L_z$.
}
\label{fig:figure1}
\end{figure*}

Figure~\ref{fig:figure1} shows instantaneous snapshots of the
velocity field,
illustrating the characteristic flow structures through the
components $\tilde{u}_x$ and $\tilde{u}_z$ and their vertical
gradients in the plane $\tilde{y}=L_y/2$.  All  simulations with $L_x=L_y=3$ and $L_z=1$ exhi\-bit one ascending and one descending plume, accompanied by two vortices.  

\begin{figure*}
\centering
\includegraphics[width=1.8\columnwidth,angle=0]{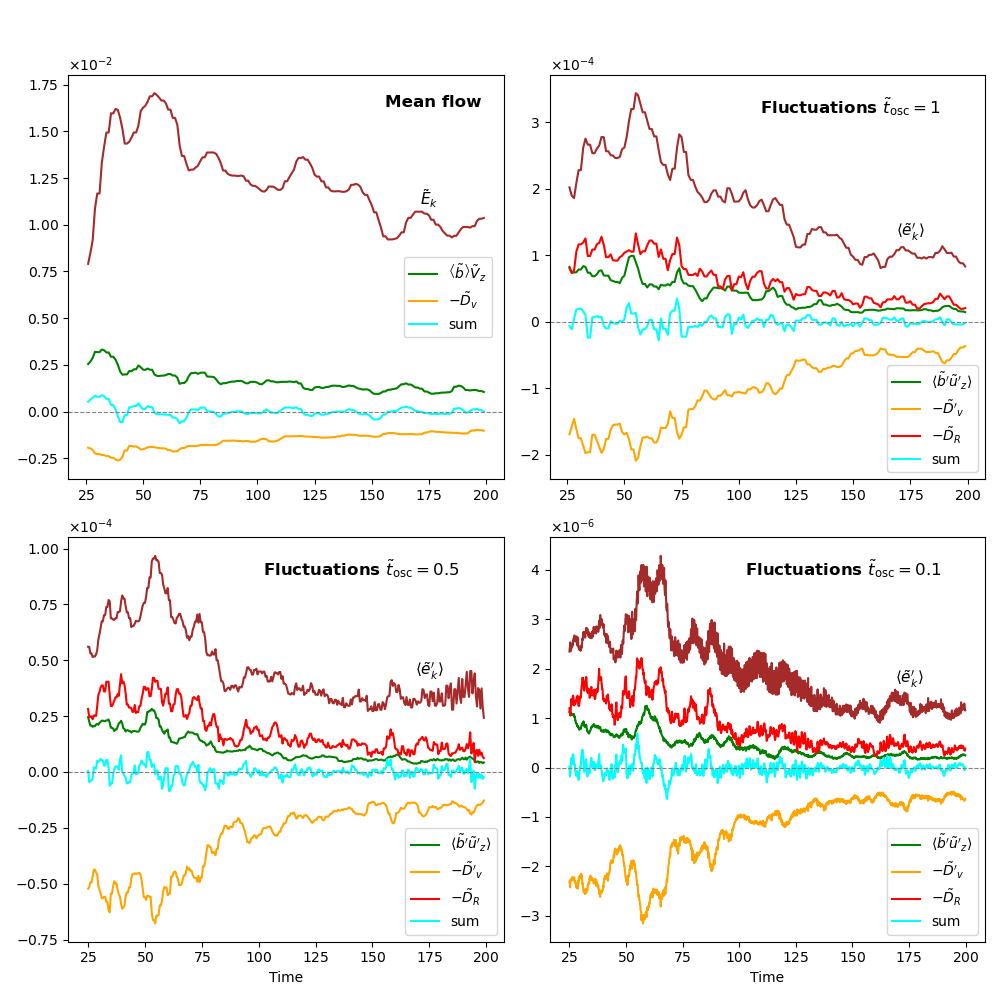}
\caption{
  Time evolution of the volume-averaged total energy budget for unforced convection at $Ra=10^6$. 
  \textbf{Top-left}: mean flow  (identical for all $\tilde{t}_{\rm osc}$). Brown: $ \tilde{E}_k  $ (mean kinetic energy);
  green: $\langle \tilde{b} \rangle  \,\tilde{V}_z$ (buoyancy work);
  orange: $-\tilde{D}_v$ (viscous dissipation);
  cyan: sum of the last two terms.
  \textbf{Other panels}: fluctuations 
  obtained using averaging windows $\tilde{t}_{\rm osc}=1$, $0.5$, $0.1$.
  Brown: $\langle \tilde{e}'_k \rangle $ (fluctuation kinetic energy);
  green: $\langle \tilde{b}' \,\tilde{u}'_z\rangle$ (buoyancy work);
  orange: $-\tilde{D}'_v$ (viscous dissipation);
  red: $-\tilde{D}_R$ (net kinetic energy assigned to the fluctuation field);
  cyan: sum of the last three terms.
  \emph{Note}: Kinetic energy is shown as energy, other terms are rates.
  All quantities are dimensionless and share the same axis for visual comparison (despite different physical dimensions). \\
 With $\tilde{t}_{\rm conv} \sim 10$, the chosen $\tilde{t}_{\rm osc}$ values filter motions well below the dominant convective timescale. The clear $t_{\rm osc}^2$ scaling of fluctuation kinetic energy confirms that the fluctuations are dominated by large-scale temporal variability rather than a forward energy cascade to small scales.
}
\label{fig:figure2}
\end{figure*}

Figure~\ref{fig:figure2} shows the total energy budget, with all terms expressed as volume-averaged quantities (i.e. integrated over the domain and divided by the total volume).

To provide a baseline against which to compare  simulations with external forcing, we 
apply the Reynolds decomposition introduced in Section~\ref{sec:timescales}, 
using a temporal block average over non-overlapping windows of duration $t_{\rm osc}$, as described in Section~\ref{sec:averaging}. 
In the absence of forcing, $t_{\rm osc}$ acts as a temporal filter that separates 
the flow into a slowly varying mean component $\mathbf{V}(t)$ and 
a residual field $\mathbf{u}' = \mathbf{u} - \mathbf{V}$, which includes both  motions 
on timescales shorter than $t_{\rm osc}$ and contributions from the temporal evolution of the mean flow within each averaging window.
Since $|\mathbf{u}'| \ll |\mathbf{u}|$ in all cases considered here, the mean flow $\mathbf{V}$ is nearly identical to the full velocity field $\mathbf{u}$, and the top-left panel of
Figure~\ref{fig:figure2} is therefore essentially identical for all values of $t_{\rm osc}$.

\subsubsection{Mean flow}

The  mean-flow energy budget (top-left) satisfies global energy conservation in the statistically steady state, as expressed by equation~(\ref{eq:globalenergymean}).  The surface flux term is negligible: energy is dissipated locally by viscosity. The cyan curve (buoyancy work minus viscous dissipation) remains near zero, confirming this balance. The  term $\tilde{D}_R$ is negative but small compared to buoyancy work and may be neglected in the mean-flow budget. The time derivative of the mean kinetic energy is also negligible, indicating that no significant energy accumulation occurs over the integration time. Equation~(\ref{eq:globalenergymean}) therefore holds without a storage term.

Surface temperature perturbations are larger at early times and decrease
as the system evolves towards a statistically steady state, with the maximum value of $\left| \tilde{b} \right|$ at the surface dropping from   0.2 at $\tilde{t}=30$ to  0.06 at $\tilde{t}=200$. 
As a result, convection is more vigorous at early times, since positive surface temperature perturbations $\delta T$  enhance  the radiative flux as $ \propto \left( T_2 + \delta T \right)^4 $.   In contrast, the rms value of $\tilde{b}$ within the flow volume increases from 0.32 to 0.48 over the course of the simu\-lation.   This behaviour reflects the fact
that, as convection becomes established, thermal fluctuations are
redistributed from the boundary into the interior, reducing surface temperature 
contrasts while amplifying temperature variations within the bulk of
the flow.
To isolate the role of the radiative boundary condition, we performed an otherwise identical simu\-lation in which the upper surface is held at a fixed radiative flux rather than obeying a blackbody condition.  In this case, surface temperature perturbations remain large throughout the simulation, with a maximum  $\left| \tilde{b} \right| \simeq 0.5$ at all times, while temperature fluctuations in the bulk are weaker, with an rms value of 0.29.  This contrast demonstrates that the blackbody radiative boundary condition promotes the redistribution of thermal variability from the surface into the interior of the flow.  Similar behaviour is obtained for both free and rigid upper boundaries.

The dimensionless convective timescale $\tilde{t}_{\rm conv} = L_z / \tilde{u}_{z,{\rm rms}}$ (see eq.~[\ref{eq:tconv}]) increases  from approximately 9 at $\tilde{t}=30$ to about 13 at $\tilde{t}=200$.   The dimensionless crossing time $\tilde{t}_{\rm cross}$ (eq.~[\ref{eq:tcross}]) remains  of order unity throughout.  The rms values of the velocity components satisfy
$\tilde{u}_{x,{\rm rms}} \simeq \tilde{u}_{y,{\rm rms}} \simeq \tilde{u}_{z,{\rm rms}} \simeq 0.1$,
while the rms values of the velocity gradients
$\partial \tilde{u}_x / \partial \tilde{z}$ and $\partial \tilde{u}_z / \partial \tilde{x}$ are
approximately 0.5, with comparable values obtained by interchanging $x$
and $y$. These scalings imply a characteristic dimensionless convective length scale
$\tilde{\lambda}_{\rm conv} \sim 0.2$.

All simulations presented here use a free upper surface.  The maximum surface  displacement satisfies $\left| \tilde{\eta} \right| \lesssim 0.01$ at all times, validating the assumption that  surface deformations remain small compared to $L_z$.   We have additionally verified that the quantities shown in Figure~\ref{fig:figure2} depend  only weakly on whether a free or rigid upper boundary is used.

\subsubsection{Fluctuations}
\label{sec:flucnoforcing}

The fluctuation budgets (other panels) show the energy of the motions on the filtered scale set by $\tilde{t}_{\rm osc}$.
Thus, $\tilde{t}_{\rm osc}=1$ retains all variability faster than unity but filters slower motions. Correspondingly, the fluctuation kinetic energy is much smaller than the mean-flow energy, indicating that most of the energy resides in motions with timescales longer than 1, consistent with $\tilde{t}_{\rm conv} \sim 10$.  

Because the fluctuations are defined as the residual with respect to a finite-time block average, they include not only motions produced by the turbulent cascade on timescales shorter than $t_{\rm osc}$, but also contributions from the temporal evolution of the slowly varying  flow within each averaging window.  To illustrate this, 
noting  that $t_{\rm osc}$ is small compared to the timescale on which the convective flow varies,
we expand the convective velocity in a Taylor series for $t$ between 
$n t_{\rm osc}$ and $(n+1) t_{\rm osc}$.  Denoting  $t_m = \left( n + 1/2 \right) t_{\rm osc} $,  this yields
${\bf u} (t) \simeq {\bf u}(t_m) + \left( t - t_m \right) \dot{\bf u} (t_m),$ 
where  the {\em dot} is a  time-derivative.  The linear term does not contribute to the block-averaged velocity ${\bf V}$ over the interval
$\left[ n t_{\rm osc}, (n+1) t_{\rm osc} \right]$, so that 
 the residual within the block is, to leading order,
${\bf u}' \equiv {\bf u}-{\bf V}= \left( t - t_m \right) \dot{\bf u} (t_m)$.  This contribution yields $\dot{u}^2 (t_m) t_{\rm osc}^2 /12$ to $\left< e'_k \right>$, consistent with the scaling $\langle \tilde{e}'_k \rangle \propto \tilde{t}_{\rm osc}^2$ observed in Figure~\ref{fig:figure2} and indicating dominance over the turbulent cascade.
This interpretation is further supported by an analysis of the temporal power spectrum of the convective flow, obtained via Fourier transform, which shows that the kinetic energy contained at the corresponding high frequencies is significantly smaller than the fluctuation energy measured using the block-averaging procedure. 
In other words, the fluctuations isolated by the averaging procedure do not  primarily represent energy injected at small spatial or temporal scales by a turbulent cascade, but rather energy already present in the large-scale flow that is assigned to the fluctuation field by the finite-time temporal filter.

The budgets of these fluctuations satisfy global energy conservation, as given by equation~(\ref{eq:globalenergyfluc}).  
Because the fluctuation kinetic energy is dominated by the temporal variability of the large-scale convective flow within the averaging window, rather than by a forward cascade to small spatial scales, the term $D_R$ should not be interpreted as a classical turbulent energy transfer.
Both the residual kinetic energy and the exchange term $D_R$ arise simultaneously from this variability. 
From the expression of ${\bf u}'$ above, we can write:
$$D_R= \frac{t_{\rm osc}^2}{12} \dot{u}_i (t_m)  \dot{u}_j (t_m) \frac{\partial V_i}{\partial x_j} (t_m) .$$  The observed $\tilde{D}_R<0$ reflects a systematic anti-correlation between the local acceleration $\dot{\mathbf{u}}$ and the block-mean gradients, which arises because the evolution of the flow is dominated by relaxation towards a statistically steady state.   Thus, $-D_R > 0$ exactly accounts for the portion of large-scale kinetic energy that the finite averaging window reassigns from the block-mean to the residual field.

The surface flux term is negligible, and there is no significant accumulation of fluctuation kinetic energy over time: viscous dissipation $-D'_v$ balances the sum of buoyancy work on fluctuations and $-D_R$.

\subsection{Forcing prescriptions and constraints}
\label{sec:externalforcing}

We now introduce an external forcing in equation~(\ref{eq:NSdim}).  
We consider the following dimensionless prescriptions:

\begin{equation}
\tilde{\bf f}_1 =  \tilde{f}_1  \hat{\bf z}, \quad
\tilde{f}_1 = - a  \tilde{z}  \tilde{x} \left(L_x -  \tilde{x} \right)  \tilde{y} \left(L_y -  \tilde{y} \right)  \sin \left( \frac{2 \pi \tilde{t}}{  \tilde{t}_{\rm osc}}  \right)  ,
\label{eq:forcing1}
\end{equation}

\begin{equation}
\tilde{\bf f}_2 =   \tilde{f}_2   \hat{\bf z}, \quad 
\tilde{f}_2 =  - a  \tilde{z}^2  \left(L_z -  \tilde{z} \right)  \tilde{x} \left(L_x -  \tilde{x} \right)  \tilde{y} \left(L_y -  \tilde{y} \right)   \sin \left( \frac{2 \pi \tilde{t}}{  \tilde{t}_{\rm osc}}  \right)  ,
\label{eq:forcing2}
\end{equation}

\begin{equation}
\tilde{\bf f}_3 =   \tilde{f}_1 \left( \hat{\bf x} + \hat{\bf z} \right) ,
\label{eq:forcing3}
\end{equation}

\begin{equation}
\tilde{\bf f}_4 =   - \nabla \Psi, \quad \Psi = \frac{a}{2}  \tilde{z}^2   \sin \left( \frac{2 \pi \tilde{x}}{ {L_x}}  \right) \sin \left( \frac{2 \pi \tilde{y}}{  {L_y}}  \right)   \sin \left( \frac{2 \pi \tilde{t}}{  \tilde{t}_{\rm osc}}  \right)  ,
\label{eq:forcing4}
\end{equation}

\noindent where $a$ is a constant amplitude varied across simulations.
All forcing functions satisfy the periodic boundary conditions in the horizontal directions. 
The form~(\ref{eq:forcing1})  is not intended to model tidal forcing in a dynamical sense, since it does not derive from a scalar potential. Rather, it provides a delibera\-tely simple test case in which the only similarity with tides is the decay of the forcing amplitude with depth. Because it directly excites only vertical velocity fluctuations, it offers a clean framework for isolating how correlations arise within the convective flow. The expression~(\ref{eq:forcing2}) has a similar structure but vanishes at the upper boundary, allowing us to assess the sensitivity of the results to near-surface forcing.  Case~(\ref{eq:forcing3}) extends~(\ref{eq:forcing1}) by including a horizontal component. This configuration is particularly useful for comparing simulations with free and rigid upper boundaries and for assessing whether horizontal forcing modifies the resulting correlations. Finally, the prescription~(\ref{eq:forcing4}) derives explicitly from a scalar potential and decreases with depth, and therefore provides the most physically faithful representation of tidal forcing considered in this study.

All forcings vary  on length scales comparable to the domain size,
corresponding to a characteristic dimensionless oscillation length scale
$\tilde{\lambda}_{\rm osc} \sim 0.5$–$1$, several times larger than the convective length scale $\tilde{\lambda}_{\rm conv}$. This ordering is consistent with expectations for tidal oscillations
in the convective envelopes of stars and giant planets. 

 
 To ensure reliable measurements of $D_R$, the forcing amplitude $a$ must satisfy the following conditions:
 
 \begin{itemize}
 
\item[(i)] As discussed in section~\ref{sec:momentum}, the external forcing must not
modify the leading-order balance of the mean flow.
This requirement is satisfied if the  mean
buoyancy  $\left< \tilde{b} \right>$ remains large compared to the Reynolds stress term $\left<  \tilde{u}'_j  \left( \partial  \tilde{u}'_i    / \partial \tilde{x}_j  \right) \right>$. 
 Multiplying both terms by $\tilde{V}_z$, this condition may be written as 
 $\left< \tilde{b} \right> \tilde{V}_z \gg \left| \tilde{D}_R \right| \tilde{\lambda}_{\rm conv} / \tilde{\lambda}_{\rm osc}$, where the factor $\tilde{\lambda}_{\rm conv}/\tilde{\lambda}_{\rm osc}$ accounts for the
difference between the characteristic convective and forcing length
scales.  Since $ \tilde{\lambda}_{\rm conv}$ is a few times smaller than $\tilde{\lambda}_{\rm osc}$, it is sufficient to require that:
\begin{equation}
 \left| \tilde{D}_R \right| \ll \left< \tilde{b} \right> \tilde{V}_z ,
 \label{eq:conditionii}
 \end{equation}
which places the system safely within the desired asymptotic regime and
therefore provides an upper bound on the forcing amplitude $a$. \\
 

\item[(ii)]   A lower limit on $a$ is set by  viscous dissipation.  For the oscillatory fluctuations to sustain Reynolds stress correlations  $\langle u'_i u'_j \rangle \partial V_i / \partial x_j$ that yield significant $D_R$, advection of fluid elements by the oscillations across the scale of the mean velocity gradients must outpace  viscous damping of the  fluctuations.  This requires the fluctuating Reynolds number:
\begin{equation}
Re' \equiv \frac{u' \lambda_{\rm conv}}{\nu} = \frac{\tilde{u}'  \tilde{\lambda}_{\rm conv}}{\tilde{\nu}}  \gtrsim 1\text{--}10  ,
\label{eq:Reynoldsfluc}
\end{equation} 
where $u'$ is the rms fluctuation velocity  and $\lambda_{\rm conv}$ is the typical length scale over
which the mean flow varies.  For $Ra=10^6$ and $Pr=1$, ${\tilde{\nu}} = 10^{-3}$.  With $\tilde{\lambda}_{\rm conv} \simeq 0.2$, this criterion implies
$\tilde{u}' \gtrsim \text{a few times \,}  5\times10^{-3}$. \\

\item[(iii)] To ensure that the block-averaged quantities isolate the externally forced oscillatory response, the forcing must produce oscillations whose kinetic
energy is significantly larger than that associated with the large-scale
variability that dominates the unforced case.

\end{itemize}

 \subsection{Simulations with external forcing}

We have performed simulations with $\tilde{t}_{\rm osc}=1$, 0.5 and 0.1, corresponding to ratios of the oscillation period to the convective
timescale in the range $0.1$--$0.01$.   Because of the constraints discussed in the previous section, namely the
lower and upper bounds on the forcing amplitude, the range of forcings that
can be explored is necessarily limited.

We find that, whenever the time- and volume-averaged fluctuation kinetic energy  $\left< \tilde{e}'_k \right>$
exceeds approximately $2 \times 10^{-3}$,    {\em the volume-averaged transfer term $\tilde{D}_R$ is positive, indicating a net transfer of kinetic energy  from the fluctuations to the mean flow.  Moreover, at fixed  fluctuation kinetic energy, the magnitude of $\tilde{D}_R$ is essentially
independent of $\tilde{t}_{\rm osc}$ over the range of oscillation periods explored. }  

In the unforced case, the fluctuation kinetic energy scales as $\left< \tilde{e}'_k \right> \propto t_{\rm osc}^2$, reflecting the temporal variability of the slowly evolving mean flow within the averaging window rather than any distinct dynamical process. In the forced case, the externally driven oscillatory response dominates the fluctuation budget, with $\left< \tilde{e}'_k \right>$  one to three orders of magnitude larger than in the unforced case at the same $t_{\rm osc}$. This large sepa\-ration in amplitude is the primary diagnostic confirming that the block-averaging procedure now isolates a physically meaningful oscillatory response rather than a filtering artefact. The sign and interpretation of $\tilde{D}_R$ are correspondingly different: whereas the negative $\tilde{D}_R$ of the unforced case reflects the finite-window accounting identity derived in Section~\ref{sec:flucnoforcing}, the positive $\tilde{D}_R$ in the forced simulations reflects a genuine physical transfer of kinetic energy from the oscillatory fluctuations to the large-scale convective flow.

Since the fluctuating kinetic energy is distributed approximately equally
among the three velocity components, we have $\tilde{u}'^{\, 2} \simeq 2 \left< \tilde{e}'_k \right> /3$.  A value $\left< \tilde{e}'_k \right> = 2 \times 10^{-3}$ therefore corresponds
to a characteristic fluctuating velocity
$\tilde{u}' \sim 0.036$, yielding a fluctuating Reynolds number
$Re' \simeq 7$.  The empirical threshold for the emergence of positive correlations is thus
consistent with the condition expressed by
equation~(\ref{eq:Reynoldsfluc}).

We interpret the results below in terms of  an effective dimensionless damping rate $\tilde{\gamma}$, defined as:
\begin{equation}
\tilde{\gamma} \equiv \frac{\tilde{D}_R }{  \tilde{u}'^{\, 2} } \simeq  \frac{3 \tilde{D}_R }{ 2 \left< \tilde{e}'_k \right>}.
\label{sec:DRscaling1}
\end{equation}

\noindent 
Although we refer to $\tilde{\gamma}$ as a damping rate, reflecting its role in depleting the fluctuation kinetic energy, it represents a transfer to the mean flow rather than a direct viscous sink.


\subsubsection{Energy budgets}

\begin{figure*}
  \centering
  \includegraphics[width=1.8\columnwidth,angle=0]{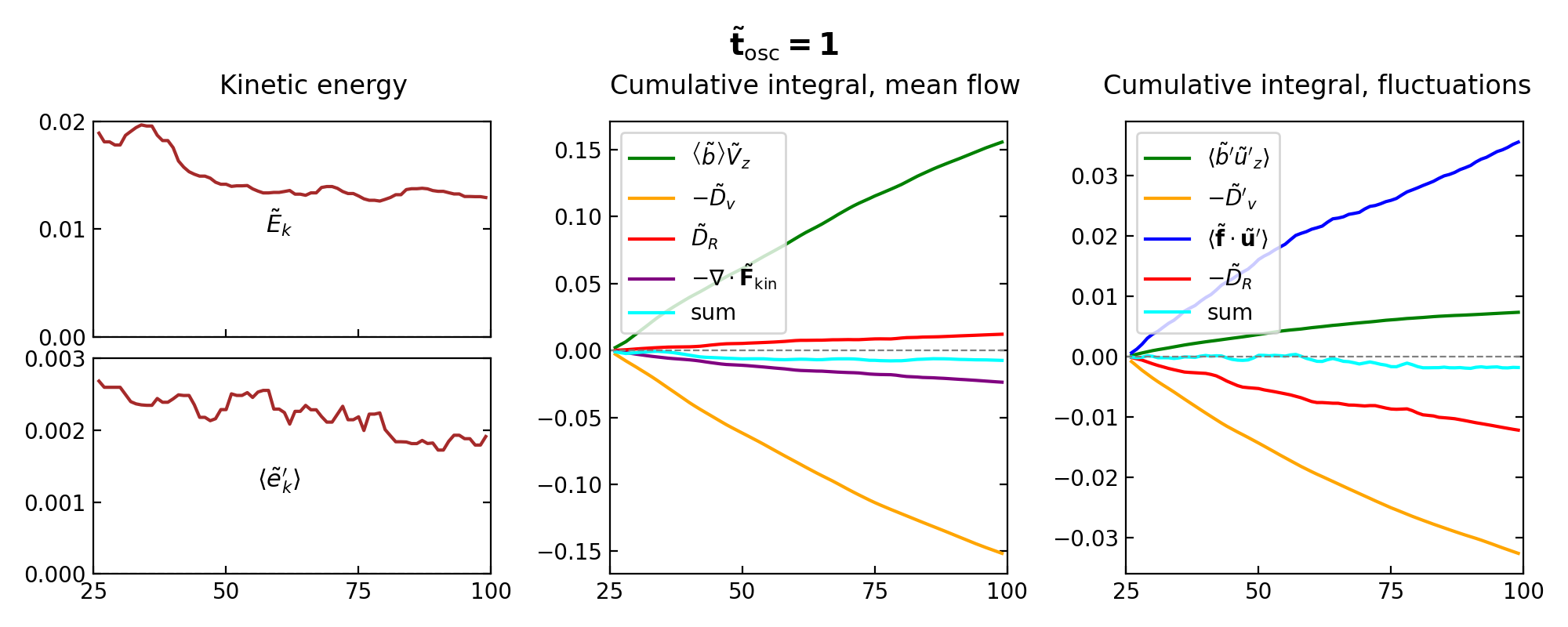}\\
  \includegraphics[width=1.8\columnwidth,angle=0]{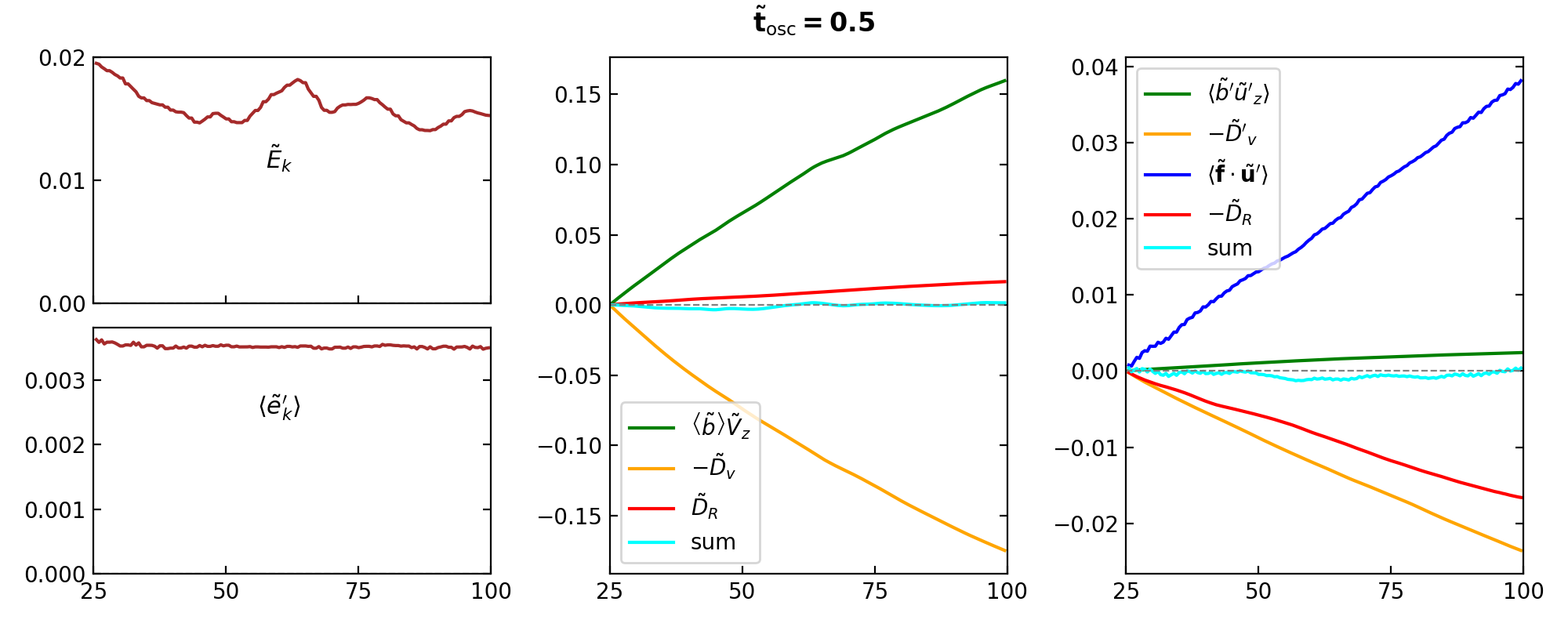}\\
  \includegraphics[width=1.8\columnwidth,angle=0]{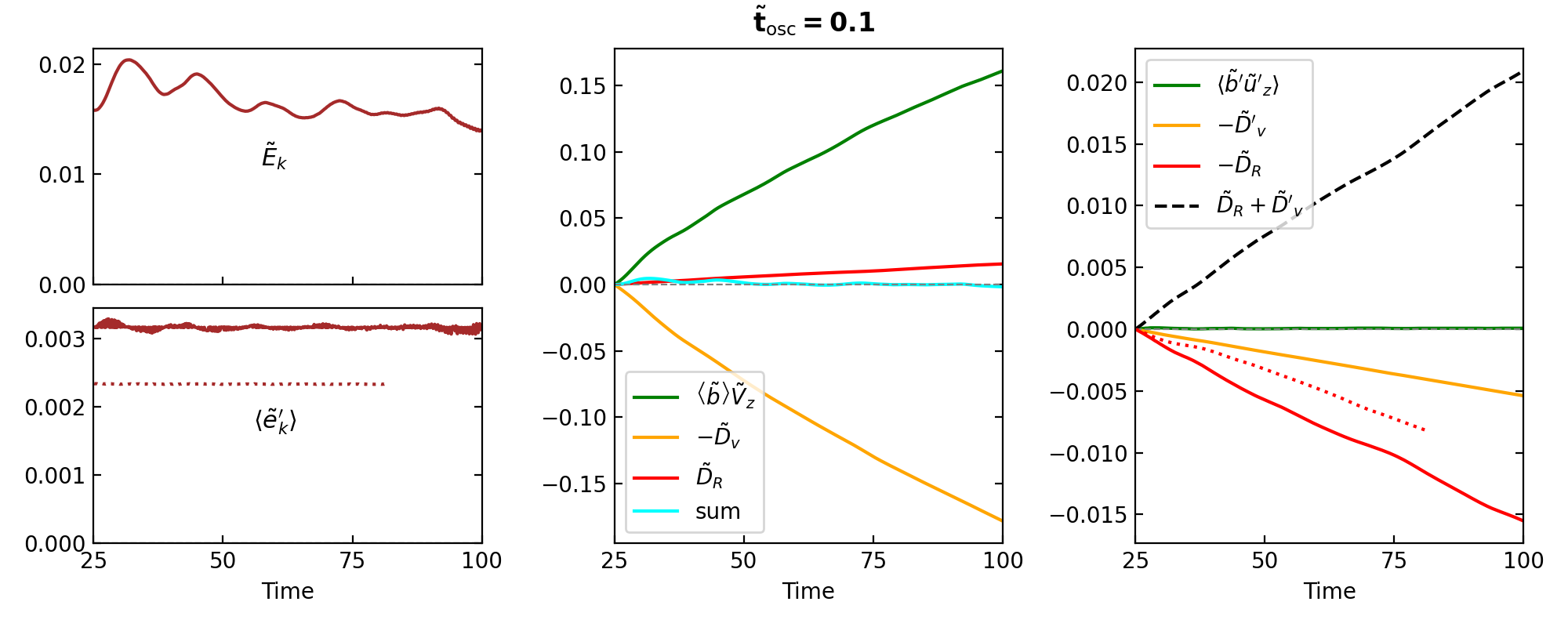}
 \caption{
  Time evolution of the volume-averaged total energy budget for forced convection with $\tilde{t}_{\rm osc}=1$ (upper row), 0.5 (middle row) and 0.1 (lower row).  The forcing is 
  $\tilde{\bf f}_1$  given by equation~(\ref{eq:forcing1}).
  The  curves correspond to the local terms appearing in the global energy equations~(\ref{eq:globalenergymean}) and~(\ref{eq:globalenergyfluc}).  Kinetic energies are volume-averaged, while all other curves show cumulative time integrals of the corresponding volume-averaged quantities.
  \textbf{Left column}: kinetic energy of the mean flow (upper panel) and of fluctuations (lower panel).
  \textbf{Middle column}: mean flow energy budget.  
  {\em Green}: $\langle \tilde{b} \rangle  \,\tilde{V}_z$ (buoyancy work);
 {\em  orange}: $-\tilde{D}_v$ (viscous dissipation);
  {\em red}: $\tilde{D}_R$ (kinetic energy {\rm gained} from the fluctuations);
  {\em purple}:  $- \nab \cdot \tilde{\bf F}_{\rm kin}$ (surface flux;  negligible for $\tilde{t}_{\rm osc}=0.5$ and 0.1);
  {\em cyan}: sum of all terms.
  \textbf{Right column}: fluctuation energy budget.
  {\em Green}: $\langle \tilde{b}' \,\tilde{u}'_z\rangle$ (buoyancy work);
  {\em orange}: $-\tilde{D}'_v$ (viscous dissipation);
  {\em red}: $-\tilde{D}_R$ (kinetic energy {\em transferred} to the mean flow);
  {\em blue}: $\left< \tilde{\bf f} \cdot \tilde{\bf u}' \right>$ (work of the external forcing); {\em cyan}: sum of all terms.  For $\tilde{t}_{\rm osc}=0.1$, the work done by the external forcing is not shown, as it is sensitive to small phase errors at short oscillation periods;
instead, we plot the cumulative time integral of $\tilde{D}’_v + \tilde{D}_R$ (black dashed curve) which, by the fluctuation energy budget, must balance the external work. 
The dotted curves in the $\tilde{t}_{\rm osc}=0.1$ row illustrate a case with smaller fluctuating kinetic energy.  
Overall, these budgets show that while the mean-flow energetics
remain dominated by buoyancy work, a positive and well-converged transfer
term $\tilde{D}_R$ systematically channels a fraction of the fluctuation energy
into the mean flow, with a time- and volume-averaged magnitude that is
essentially independent of $\tilde{t}_{\rm osc}$ for the cases shown.
}
\label{fig:figure3}
\end{figure*}

We now describe representative simulations at  $Ra=10^6$  that illustrate the  energy budgets of the 
mean flow and  fluctuations under external forcing.  We begin with the  purely vertical forcings $\tilde{\bf f}_1 $ and $\tilde{\bf f}_2 $, using amplitudes chosen in accordance with the constraints outlined above. Simulations employing the other forcings  ($\tilde{\bf f}_3 $ and the potential-derived $\tilde{\bf f}_4 $) are discussed in subsequent subsections.  Results for $\tilde{\bf f}_1 $ are shown in 
Figure~\ref{fig:figure3}. The viscous timescale associated with a convective length scale is $\tilde{\lambda}_{\rm conv}^2 / \tilde{\nu} \simeq 40$.  All simulations are run for at least $\tilde{t}=100$, ensuring that viscous effects are fully captured over the convective scales of interest.

Table~\ref{tab:table1} summarises the time- and volume-averaged fluctuating kinetic energy $\langle \tilde{e}'_k \rangle $, transfer rate $\tilde{D}_R$ and resulting damping rate $\tilde{\gamma}$ for all the simulations.

\begin{table}
\centering
\caption{
Time- and volume-averaged fluctuating kinetic energy $\langle \tilde{e}'_k \rangle $, transfer rate $\tilde{D}_R$ and resulting damping rate $\tilde{\gamma}$ for the simulations shown in
Figure~\ref{fig:figure3} to~\ref{fig:figure8}.
}
\label{tab:table1}
\begin{tabular}{cccccc}
\hline
forcing &
$\tilde{t}_{\rm osc}$  &
$\langle \tilde{e}'_k \rangle$ &
$\tilde{D}_R$ &
$\tilde{\gamma}$ & Figures \\
\hline
${\bf f}_1$ &
1.0 
& $2.2\times10^{-3}$ 
& $1.6\times10^{-4}$ & 0.1 & \ref{fig:figure3} and \ref{fig:figure5}  \\
--- &
0.5 
& $3.5\times10^{-3}$ 
& $2.2\times10^{-4}$ & 0.1 & --- \\
--- &
0.1 
& $3.2\times10^{-3}$ 
& $2.1\times10^{-4}$ & 0.1 & --- \\
--- &
---
& $2.3\times10^{-3}$  
& $1.5\times10^{-4}$ & 0.1 & --- \\
${\bf f}_2$ &
1.0 
& $2.4\times10^{-3}$ 
& $3.0\times10^{-4}$ & 0.19 & \ref{fig:figure4} \\
${\bf f}_3$ &
1.0 
& $2.3\times10^{-3}$ 
& $1.9\times10^{-4}$ & 0.1 & \ref{fig:figure6} \\
${\bf f}_4 = - \nab \Psi$ &
0.5 
& $1.8\times10^{-3}$ 
& $2.2\times10^{-4}$ & 0.19 & \ref{fig:figure7} and \ref{fig:figure8}  \\
\hline
\end{tabular}
\end{table}

The left column of Figure~\ref{fig:figure3} shows the time evolution of the volume-averaged kinetic energies of the mean flow and fluctua\-tions.
The panels in the middle and right columns   show the cumulative time integrals, from $\tilde{t}=25$, of the terms appearing in the global energy equations~(\ref{eq:globalenergymean}) and~(\ref{eq:globalenergyfluc}), after normalisation by the volume.
In the fluctuation energy budget, energy input is dominated by the external
forcing, while the contribution from fluctuating buoyancy is small for $\tilde{t}_{\rm osc}=1$ and negligible at shorter oscillation timescales.  
The fraction of  the fluctuation energy  transferred  to the mean flow via the $\tilde{D}_R$ term
 is approximately 30\%, 40\% and 74\% for $\tilde{t}_{\rm osc}=1$, 0.5 and 0.1, respectively, with the  remainder dissipated viscously.  
 
For $\tilde{t}_{\rm osc}=1$ and 0.5, the residual of the fluctuation ener\-gy budget (cyan curves) is one to two orders of magnitude smaller than $\tilde{D}_R$,  indicating excellent energy conservation.  No energy flux through the upper surface is included in the fluctuation budget, confirming that such fluxes are negligible in all cases considered.

For $\tilde{t}_{\rm osc}=1$ and 0.5,  the time- and volume-averaged external work $\left< \tilde{\bf f} \cdot \tilde{\bf u}' \right>$ is 
$4.7 \times 10^{-4}$ and $5.1 \times 10^{-4}$, respectively.
For $\tilde{t}_{\rm osc}=0.1$, however, the net work becomes highly sensitive to extremely small phase differences between $\tilde{\bf f}$ and $\tilde{\bf u}'$.  At such short periods, tiny numerical phase offsets accumulate over long integrations and dominate the inferred time average.  We therefore do not rely on a direct evaluation of the forcing work in this case.
Crucially, the  transfer term $\tilde{D}_R$ itself remains robust.
Its value is insensitive to spatial resolution and to the procedure used to estimate the forcing work.
Furthermore, for $\tilde{t}_{\rm osc}=0.1$, over $25 \le \tilde{t} \le 100$, the time-averaged residual of the mean-flow energy budget is more than an order of magnitude smaller than $\tilde{D}_R$,
confirming that energy conservation is satisfied to high precision. The greater robustness of $\tilde{D}_R$, compared with the direct evaluation of the
external work, arises because 
the latter depends on small phase offsets between $ {\bf f} $ and $ {\bf u}'$ which  are highly sensitive to numerical errors.  By contrast, as will be shown below,  $\tilde{D}_R$ results from spatial correlations  controlled by the large-scale convective dynamics and is therefore much less sensitive to discretisation and time-stepping choices.  This point is  discussed further in Section~\ref{sec:discussion}.  

\noindent The mean-flow energy budget is dominated by buoyancy work, whose time- and volume-averaged value is approximately $2 \times 10^{-3}$ for all values of $\tilde{t}_{\rm osc}$ considered here. The ener\-gy transfer from the fluctuations via the $\tilde{D}_R$ term accounts for about 10\% of the total.
Most of the mean-flow energy is dissipated  viscously.  For $\tilde{t}_{\rm osc}=1$, however, there  is also a significant contribution from kinetic energy flux through the upper surface.   In this case, essentially all of the energy gained from $D_R$, and slightly more, is transported to the surface by the kinetic component of the energy flux, $\tilde{\bf F}_{\rm kin}$, defined in equation~(\ref{eq:kineticflux}).  Contributions to the surface flux arising from pressure and viscous stresses are not shown,  as they are significantly smaller. 
 
\noindent The fact that  $\tilde{D}_R$ remains more than an order of magnitude
smaller than the buoyancy work acting on the mean flow confirms
that the condition~(\ref{eq:conditionii})  is satisfied in a volume-averaged sense.

Among the three cases considered, the simulation with
$\tilde{t}_{\rm osc}=1$ exhibits the largest level of noise and the
largest resi\-duals in the mean-flow energy budget.
This occurs because, in this case, the timescale of the externally driven fluctuations is
comparable to that of the fastest convective plumes. Although such plumes are rare, this partial  overlap makes the separation between
mean flow and fluctuations less clean at the diagnostic level.
Consistent with this interpretation, a substantial fraction of the
energy transferred from the fluctuations to the mean flow is transported
to the surface in this case.

\begin{figure*}
  \centering
  \includegraphics[width=1.8\columnwidth,angle=0]{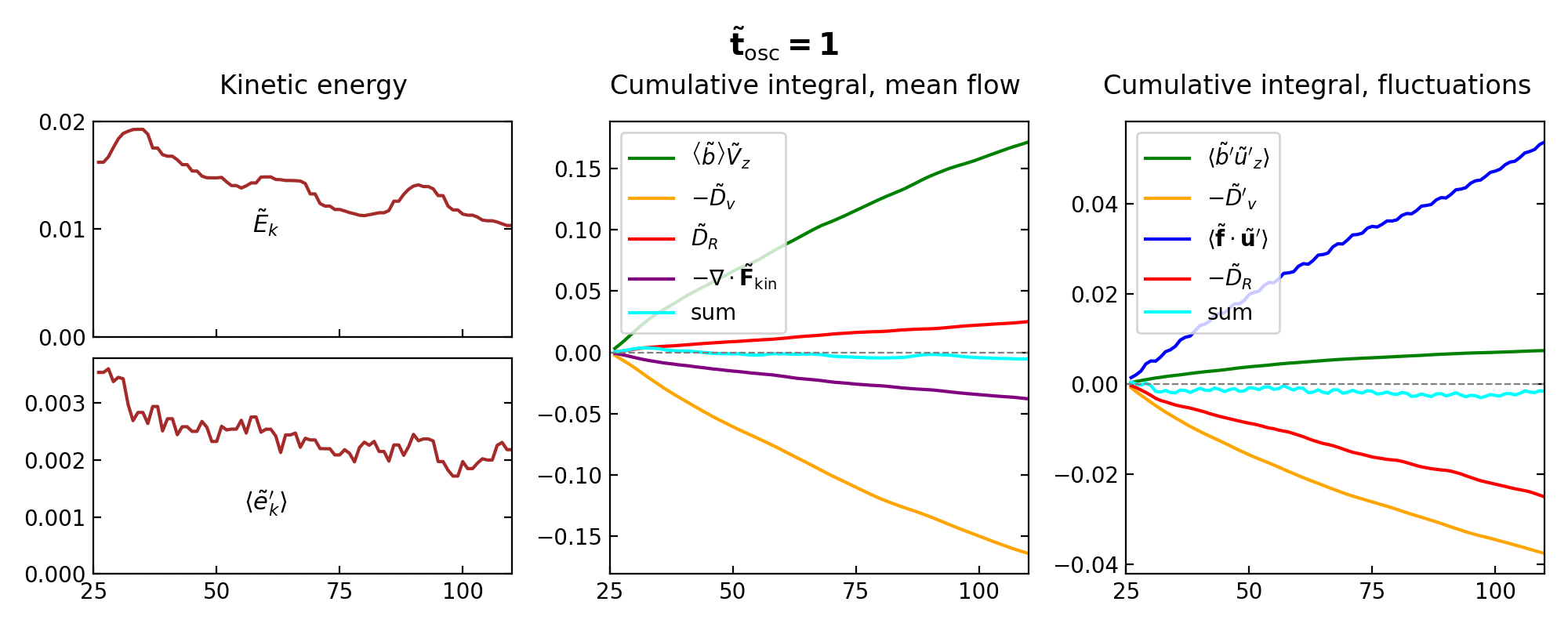}
 \caption{
 Same as Figure~\ref{fig:figure3}, but for the external forcing $\tilde{\bf f}_2$ given by equation~(\ref{eq:forcing2}) and for $\tilde{t}_{\rm osc}=1$.  The overall structure of the energy budget is very similar to that obtained with the forcing $\tilde{\bf f}_1$, although a larger fraction of the fluctuating kinetic energy is transferred to the mean flow.
  }
\label{fig:figure4}
\end{figure*}

Figure~\ref{fig:figure4} shows the energy budget in the same format as Figure~\ref{fig:figure3}, but using the external forcing $\tilde{\bf f}_2$ (eq.~[\ref{eq:forcing2}]) at $\tilde{t}_{\rm osc}=1$.  
The time- and volume-averaged values of $\left< \tilde{e}'_k \right>$ and $\tilde{D}_R$ are listed in table~\ref{tab:table1}, with  $\left< \tilde{\bf f} \cdot \tilde{\bf u}' \right> =6.3 \times 10^{-4}$.  Approximately 40\% of the fluctuation kinetic energy is transferred to the mean flow, a larger fraction  than observed with forcing $\tilde{\bf f}_1$.  Aside from this quantitative difference, the energy budgets are very similar to those found for $\tilde{\bf f}_1$, indicating that the results are not sensitive to whether the forcing vanishes at the upper boundary.

The maximum surface displacement $\left| \tilde{\eta} \right|$ remains small  in all simulations, reaching values of at most 0.06 for $\tilde{t}_{\rm osc}=1$ and 0.02 for the shorter oscillation periods $\tilde{t}_{\rm osc}=0.5$ and 0.1.

\subsubsection{Magnitude of $D'_v$ and dissipation length scale}

The rate $\tilde{D}'_v$ of fluctuating kinetic energy loss through viscous dissipation, defined in equation~(\ref{eq:Dvfluc}), may be estimated as:
\begin{equation}
\tilde{D}'_v \sim 2 \, \tilde{\nu}  \, \dfrac{\tilde{u}'^{\, 2} }{\tilde{\lambda}_{\rm dis}^2},
\end{equation}
where $\tilde{\lambda}_{\rm dis}$ is the characteristic dimensionless length scale at which the fluctuating kinetic energy is dissipated.

The nonlinear advection term ${\bf u} \cdot \nab {\bf u}$ transfers energy from the forcing scale  $\tilde{\lambda}_{\rm osc}$ to smaller spatial  scales, so that gra\-dients of ${\bf u}'$ are dominated by structures at scales $\tilde{\lambda}_{\rm dis} < \tilde{\lambda}_{\rm osc}$.  However, when the oscillation timescale $\tilde{t}_{\rm osc}$ becomes
short, the time available for nonlinear self-advection within one
oscillation period is reduced. 
As a result, the fluctuations are less efficient at generating
small-scale structures, and the cascade stops at progressively
larger dissipation scales.

Using the measured values of $\left< \tilde{e}'_k \right>$ and $\tilde{D}'_v$ shown in Figu\-re~\ref{fig:figure3},
we infer dissipation length scales $\tilde{\lambda}_{\rm dis}=0.07$, 0.1 and 0.2 for $\tilde{t}_{\rm osc}=1$, 0.5 and 0.1, respectively, in the case of the forcing given by equation~(\ref{eq:forcing1}).  

In the absence of external forcing, the dissipation length scale associated with the fluctuations, as inferred from Figu\-re~\ref{fig:figure2}, is $\tilde{\lambda}_{\rm dis} \sim 0.03$ for all values of $\tilde{t}_{\rm osc}$, comparable to the grid scale. A similar dissipation length scale is obtained for the mean flow. In this regime, the fluctuations arise  from the temporal evolution of the convective flow within each averaging window rather than from a distinct dynamical cascade. The corresponding dissipation therefore reflects the viscous damping of this residual field at the smallest resolved scales, rather than a transfer of energy across an inertial range. The dissipation length scale is thus set by the numerical resolution and is independent of $\tilde{t}_{\rm osc}$.

\subsubsection{Reynolds-stress contributions to $D_R$ and effective damping scaling}
\label{sec:DR}

\begin{figure*}
\centering
\includegraphics[width=1.4\columnwidth,angle=0]{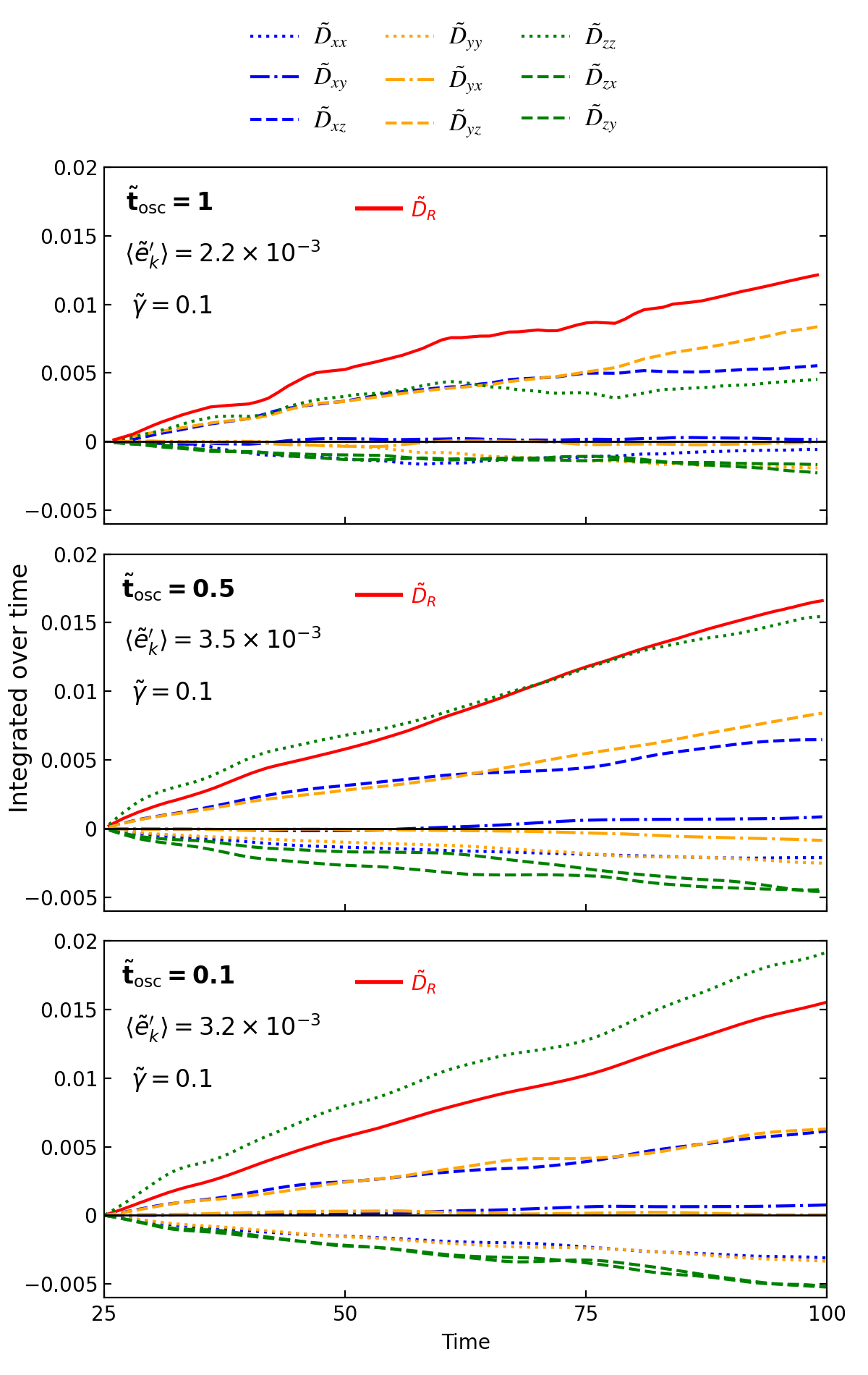}
\caption{
  Cumulative time integrals of the volume-averaged transfer terms
  $ \tilde{D}_{ij} \equiv \left< \tilde{u}'_i   \tilde{u}'_j \right> \left( \partial \tilde{V}_i / \partial \tilde{x}_j \right)$, defined with  no summation over $i$ and $j$, for the same simulations as in Figure~\ref{fig:figure3}, corresponding to the purely vertical forcing $\tilde{\bf f}_1$ given by equation~(\ref{eq:forcing1}).
  The net transfer is
    $\tilde{D}_R = \sum_{i,j} \tilde{D}_{ij}$.
  Positive values of  the volume-averaged $\tilde{D}_{ij}$ correspond to a   transfer of kinetic energy from the fluctuations to the mean flow. The upper, middle and lower panels correspond to $\tilde{t}_{\rm osc} =1$, 0.5 and 0.1, respectively.
  Dotted blue, orange  and green curves show $\tilde{D}_{xx}$, $\tilde{D}_{yy}$ and $\tilde{D}_{zz}$.
  Dashed curves represent mixed components: $\tilde{D}_{xz}$ and $\tilde{D}_{yz}$ (blue and orange), and  $\tilde{D}_{zx}$ and $\tilde{D}_{zy}$ (green).
 Dash-dotted  curves  show $\tilde{D}_{xy}$ (blue) and $\tilde{D}_{yx}$ (orange).  
 The solid red curve shows the total transfer $\tilde{D}_R$.   In each panel, the indicated value of $\left< \tilde{e}'_k \right>$ corresponds to the time- and volume-averaged  fluctuating kinetic energy over the interval shown, and $\tilde{\gamma} \equiv {\tilde{D}_R }/{  \tilde{u}'^{\, 2} } $ denotes the effective damping rate.   In all cases, the net positive volume-averaged transfer $\tilde{D}_R$ is dominated by the contributions $\tilde{D}_{zz}$, $\tilde{D}_{xz}$ and $\tilde{D}_{yz}$, indicating that correlations involving vertical velocity fluctuations and vertical gradients of the mean flow play the primary role in transferring kinetic energy from fluctuations to the mean flow across oscillation timescales. Despite the change in oscillation timescale, $\tilde{\gamma}$ remains close to 0.1 in all three cases.
}
\label{fig:figure5}
\end{figure*}

Figure~\ref{fig:figure5} shows the cumulative time integrals of the volume-averaged Reynolds-stress components $\tilde{D}_{ij} \equiv \langle \tilde{u}'_i \tilde{u}'_j \rangle (\partial \tilde{V}_i / \partial \tilde{x}_j)$, with no summation implied over $i$ and $j$, and their sum $\tilde{D}_R$, 
for the same simulations as in Figure~\ref{fig:figure3}, corresponding to the purely vertical forcing $\tilde{\bf f}_1$ given by equation~(\ref{eq:forcing1}).
The upper, middle, and lower panels correspond to $\tilde{t}_{\rm osc}=1$, 0.5 and 0.1, respectively.
In all cases, the cumulative sum of the individual contributions yields a positive volume-averaged net transfer $\tilde{D}_R$, confirming that kinetic energy is systematically transferred from the fluctuations to the mean flow.

A component-wise decomposition shows that the dominant positive contributions arise from $\tilde{D}_{zz}$, $\tilde{D}_{xz}$ and $\tilde{D}_{yz}$, indicating that correlations involving vertical velocity fluctuations and vertical gradients of the mean flow control the energy transfer across all oscillation timescales examined. 
By contrast, the components $\tilde{D}_{zx}$ and $\tilde{D}_{zy}$, associated with  horizontal shear, contribute negatively.   These terms typically offset  the contribution from $\tilde{D}_{zz}$, such that
 the net volume-averaged transfer is well approximated by $\tilde{D}_R \sim \tilde{D}_{xz} +  \tilde{D}_{yz}$.
The physical origin of this dominance is discussed in Section~\ref{sec:discussion}.

Results obtained with the purely vertical forcing $\tilde{\bf f}_2$ that vanishes at the surface, given by equation~(\ref{eq:forcing2}) (not shown), are quantitatively similar, with the same Reynolds-stress components contributing positively and negatively to the volume-averaged $\tilde{D}_R$.

For all three oscillation periods shown in Figures~\ref{fig:figure3} and~\ref{fig:figure5}, we find that the effective dimensionless damping rate ~(\ref{sec:DRscaling1}) is $\tilde{\gamma} \simeq 0.1$.
In the case where the forcing vanishes at the surface and $\tilde{t}_{\rm osc}=1$, the inferred value of $\tilde{\gamma}$ is larger, approaching $\tilde{\gamma}\simeq 0.2$.  The theoretical basis for the period-independence of $\tilde{\gamma}$ is discussed in Section~\ref{sec:discussion2}.  

\noindent Consider one of the components  $\tilde{D}_{ij} $ that contributes positively to $\tilde{D}_R$.  If the correlation between the fluctuations and the mean-flow velocity gradient were perfect, one would expect $\tilde{D}_{ij}  \sim  \tilde{u}'^{\, 2}  \left( \partial \tilde{V}_i / \partial \tilde{x}_j \right)_{\rm rms}$.  In our simulations,  the rms value of the mean-flow velocity gradient  is approximately 0.5, which  would imply a  volume-averaged value of $\tilde{D}_R$ several  times larger than that observed.  

The measured values therefore indicate that the correlations are not
perfect, but remain of order unity.  The observed magnitude of the volume-averaged $\tilde{D}_R$ is instead consistent with replacing $\left( \partial \tilde{V}_i / \partial \tilde{x}_j \right)_{\rm rms}$ by the characteristic
mean-flow shear $\tilde{V}_{\rm rms} / L_z = 1/\tilde{t}_{\rm conv}.$ This leads to the scaling:
\begin{equation}
{D}_R   \sim  \dfrac{ u'^{\, 2} }{t_{\rm conv}},
\end{equation}
which reflects strong correlations between the
fluctuations and the large-scale mean flow.  In terms of the effective damping rate, this scaling implies:
\begin{equation}
\gamma   \sim  \dfrac{1}{t_{\rm conv}}.
\end{equation}

\subsubsection{The effect of changing the Rayleigh number}
\label{sec:Rayleighnumber}

We  investigate the effect of varying the Rayleigh number $Ra$ while keeping the Prandtl number $Pr$ fixed. Decreasing $Ra$ increases both the viscosity and the thermal diffusivity, as implied by equations~(\ref{eq:kappa}) and~(\ref{eq:nu}), thereby weakening buoyant driving relative to viscous and thermal diffusion.  

For the reference case shown in Figures~\ref{fig:figure3} and~\ref{fig:figure5} with $\tilde{t}_{\rm osc}=1$ and $Ra=10^6$, corresponding to the purely vertical forcing ${\bf f}_1$, we find that, when $Ra$ is reduced to $10^5$, the volume-averaged contributions $D_{xz}$, $D_{yz}$, and $D_{zz}$ are initially positive but decay significantly after a time comparable to the viscous
timescale associated with a convective length scale,
$\tilde{\lambda}_{\rm conv}^2/\tilde{\nu}$.

For the alternative  purely vertical forcing ${\bf f}_2$, correspon\-ding to the case shown in Figure~\ref{fig:figure4} at $Ra=10^6$, we find that when $Ra$ is
reduced to $10^5$ the volume-integrated transfer $\tilde{D}_R$ remains comparable
to its value at $Ra=10^6$.  However, its decomposition is significantly altered:
the contribution from $D_{zz}$ increases, while the relative contribution from
horizontal shear, quantified by $(D_{xz}+D_{yz})/D_R$, decreases accordingly.

\subsubsection{Rigid upper surface}
\label{sec:rigid}

We now examine the effect of replacing the free upper surface by a rigid
boundary.
In this case, the boundary conditions at $\tilde{z}=L_z$ (see section~\ref{sec:freesurface}) are changed to $\tilde{u}_z=0$ and
$
   {\partial \tilde{b}}/{\partial \tilde{z}} = 1  -  \left( 1 +   {\tilde{b}}/{\tilde{T}_2}\right)^4 .
$
The normal stress condition~(\ref{eq:sigmazz}) is replaced by a pressure constraint, enforced
using a standard $\tau$-formulation \citep{Burns2020}.

For a rigid upper surface, simulations with purely vertical forcing do not
yield a positive volume-averaged transfer term $\tilde{D}_R$ over the duration of the simulations, which extend up to  $\tilde{t} \sim 500$.
A positive $\tilde{D}_R$ is obtained only when the forcing includes a
horizontal component.  Accordingly, we consider the forcing
${\bf f}_3 $  defined in equation~(\ref{eq:forcing3}).
In this configuration, the impermeability condition $\tilde{u}_z=0$ at the
upper boundary requires pressure forces to locally balance both buoyancy and
the imposed vertical forcing near the surface, thereby strongly modifying the
structure of the fluctuating response.

\begin{figure*}
\centering
\includegraphics[width=1.4\columnwidth,angle=0]{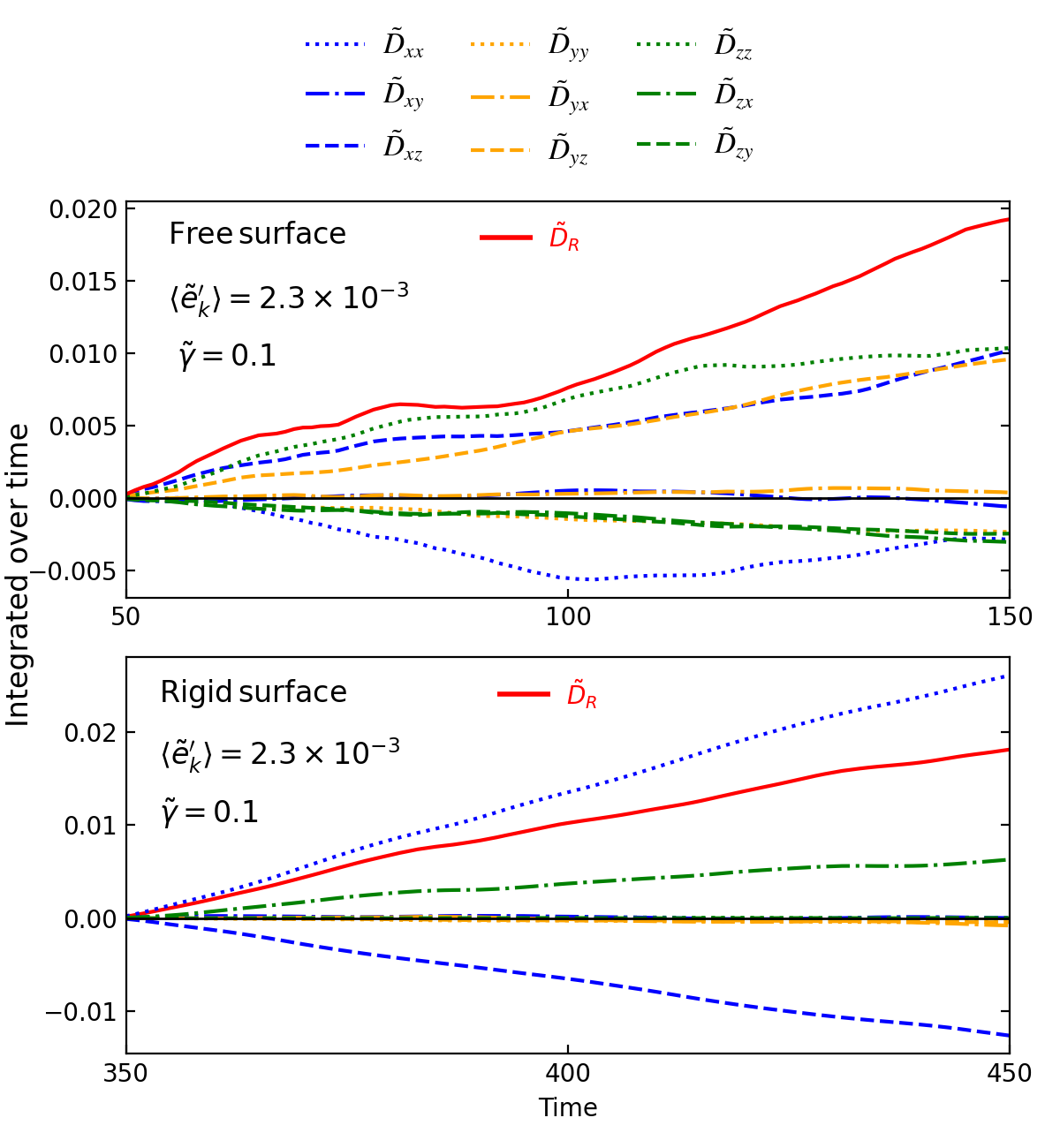}
\caption{
  Same as Figure~\ref{fig:figure5}, but for the external forcing  ${\bf f}_3$ given by equation~(\ref{eq:forcing3}) with $\tilde{t}_{\rm osc}=1$, comparing  
 a free upper surface (upper panel) and a rigid one (lower panel).  In contrast to Figure~\ref{fig:figure5},  the inclusion of a horizontal forcing component  breaks the symmetry between the mixed vertical terms.  As a result, $\tilde{D}_{zx}$ (dash-dotted green curve) and $\tilde{D}_{zy}$ (dashed green curve) are shown separately.
The indicated values of $\langle \tilde{e}'_k \rangle$ denote the time- and volume-averaged fluctuating kinetic energy over the interval displayed. 
For a free upper surface, the structure of the transfer terms remains qualitatively similar to that obtained with purely vertical forcing.  For a rigid upper surface subjected to the same mixed forcing, the behaviour differs markedly from the free-surface case, with correlations associated with the horizontal forcing dominating the net transfer.  Note also that simulations with a rigid surface and purely vertical forcing do not yield a positive volume-averaged  $\tilde{D}_R$.
}
\label{fig:figure6}
\end{figure*}

Figure~\ref{fig:figure6} shows the  cumulative time  integrals of the
Reynolds-stress transfer terms, in the same format as
Figu\-re~\ref{fig:figure5}, for this mixed forcing ${\bf f}_3 $ and for both a free and a
rigid upper surface.
In the free-surface case (upper panel), the structure of the transfer terms is
qualitatively similar to that obtained with purely vertical forcing:
the dominant positive contributions arise from
$\tilde{D}_{zz}$, $\tilde{D}_{xz}$ and $\tilde{D}_{yz}$,
yielding an effective damping rate
$\tilde{\gamma} \simeq 0.1$.

For a rigid upper surface (lower panel), the emergence of a positive  $\tilde{D}_R$ occurs only after a substantially longer transient.
Moreover, the structure of the correlations responsible for the energy  transfer is
fundamentally different from that obtained in the free-surface case.
Only the components $\tilde{D}_{xx}$ and $\tilde{D}_{zx}$ contribute
positively to the volume-averaged $\tilde{D}_R$, while the contributions
associated with vertical velocity fluctuations and vertical gradients of the mean flow are
strongly reduced, with  $\tilde{D}_{xz}$ becoming negative.
This behaviour, together with the absence of positive $\tilde{D}_R$ under purely vertical forcing, indicates that in the rigid-surface case the net energy transfer is dominated by correlations associated with the horizontally forced response, rather than by the vertical shear and buoyancy-driven interactions that dominate in the free-surface case.
We have also verified that the same qualitative behaviour is obtained when ${\bf f}_3=f_1(\hat{\bf x}+\hat{\bf z})$ is replaced by $f_2(\hat{\bf x}+\hat{\bf z})$, confirming that the
results are not sensitive to the detailed vertical structure of the forcing.

These trends are consistent with additional simulations employing purely horizontal forcing in the $x$-direction (not shown). 
For a rigid upper surface, 
$\tilde{D}_{zx}$ contributes positively while $\tilde{D}_{xz}$ is negative, whereas the signs of these two contributions are reversed in the free-surface case.


\subsubsection{Forcing deriving from a potential}
\label{sec:potential}
 
The forcing  ${\bf f}_4 = - \nab \Psi$,  derived from a scalar potential  $\Psi$ given by equation~(\ref{eq:forcing4}), mimics tidal forcing.  Using incompressibi\-lity ($\nabla \cdot \tilde{\bf u}’ = 0$), the net work done by the forcing  on the flow can be rewritten as:
\begin{equation}
 \int   \left< \tilde{\bf f} \cdot \tilde{\bf u}' \right> {\rm d} v = - \int    \nab  \cdot  \left<  \Psi  \tilde{\bf u}'  \right> {\rm d} v = - \int \left<  \Psi  \tilde{\bf u}'  \right> \cdot \hat{\bf n} \; {\rm d} s ,
 \label{eq:workpotential}
\end{equation}
\noindent where the surface integral  is taken over the domain bounda\-ries  and  $\hat{\bf n}$ is the outward unit normal. For rigid boundaries ($\tilde{\bf u}’ \cdot \hat{\bf n} = 0$ everywhere), the surface term vanishes, so the for\-cing performs no net work on the flow.  Kinetic energy is injected locally in some regions and removed in others, but these contributions cancel in the volume integral.  With a free surface present, however, the boundary term is generally non-zero, allowing the forcing to perform net work  through the surface. In this case, the potential forcing  acts effectively like a pressure  perturbation applied at the boundary,  enabling energy  exchange between the exterior and the fluid domain.

\begin{figure*}
  \centering
  \includegraphics[width=1.8\columnwidth,angle=0]{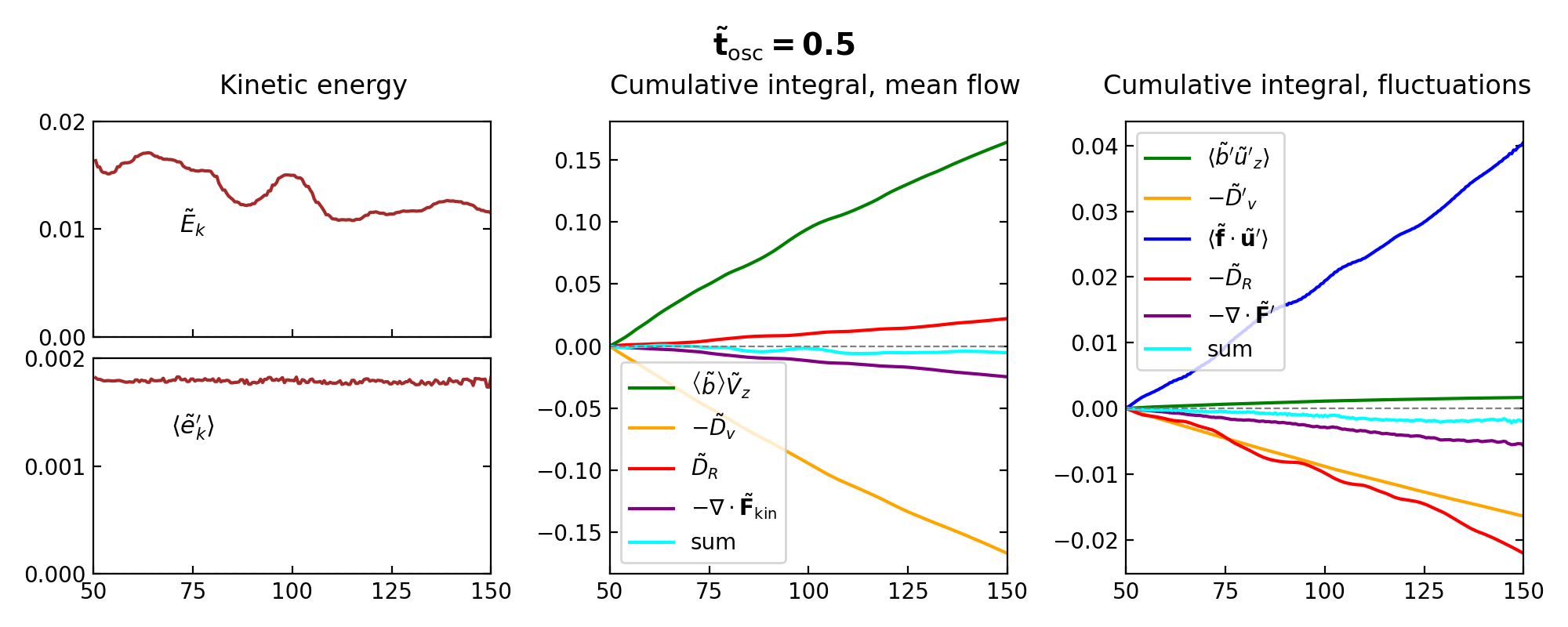}
 \caption{Energy budget for the potential-derived forcing 
$\tilde{\bf f}_4 = -\nabla \Psi$ (equation~\ref{eq:forcing4}) 
at $\tilde{t}_{\rm osc}=0.5$, shown in the same format as 
Figure~\ref{fig:figure3}. 
Both  the mean-flow and fluctuation budgets include a significant surface-flux contribution, 
reflecting  net work transmission through the boundary. 
  }
\label{fig:figure7}
\end{figure*}

Figure~\ref{fig:figure7} shows the energy budget  for  $\tilde{\bf f}_4$ at $\tilde{t}_{\rm osc}=0.5$, in the same format as Figure~\ref{fig:figure3}.  
The time- and volume-averaged values of $\left< \tilde{e}'_k \right>$ and $\tilde{D}_R$ are listed in table~\ref{tab:table1}, with  $\left< \tilde{\bf f} \cdot \tilde{\bf u}' \right> =4.1 \times 10^{-4}$.  Approximately 50\% of the fluctuating kinetic energy is transferred to the mean flow, a larger fraction  than in the cases with forcings $\tilde{\bf f}_1$ and $\tilde{\bf f}_2$.  Unlike the purely vertical forcing case, both the mean-flow and fluctuation budgets now exhibit a significant surface-flux contribution. For the mean flow, this is dominated by the kinetic component of the energy flux, $\tilde{\bf F}_{\rm kin}$, whereas for the fluctuations the dominant contribution arises from pressure fluctuations, $\left< \tilde{P}' \tilde{u}'_j \right>$ in equation~(\ref{eq:flucflux}).  The residual in both the mean-flow and fluctuation budgets is   an order of magnitude  smaller than $\tilde{D}_R$, confirming that the transfer term satisfies energy conservation to high precision.

\begin{figure*}
\centering
\includegraphics[width=1.4\columnwidth,angle=0]{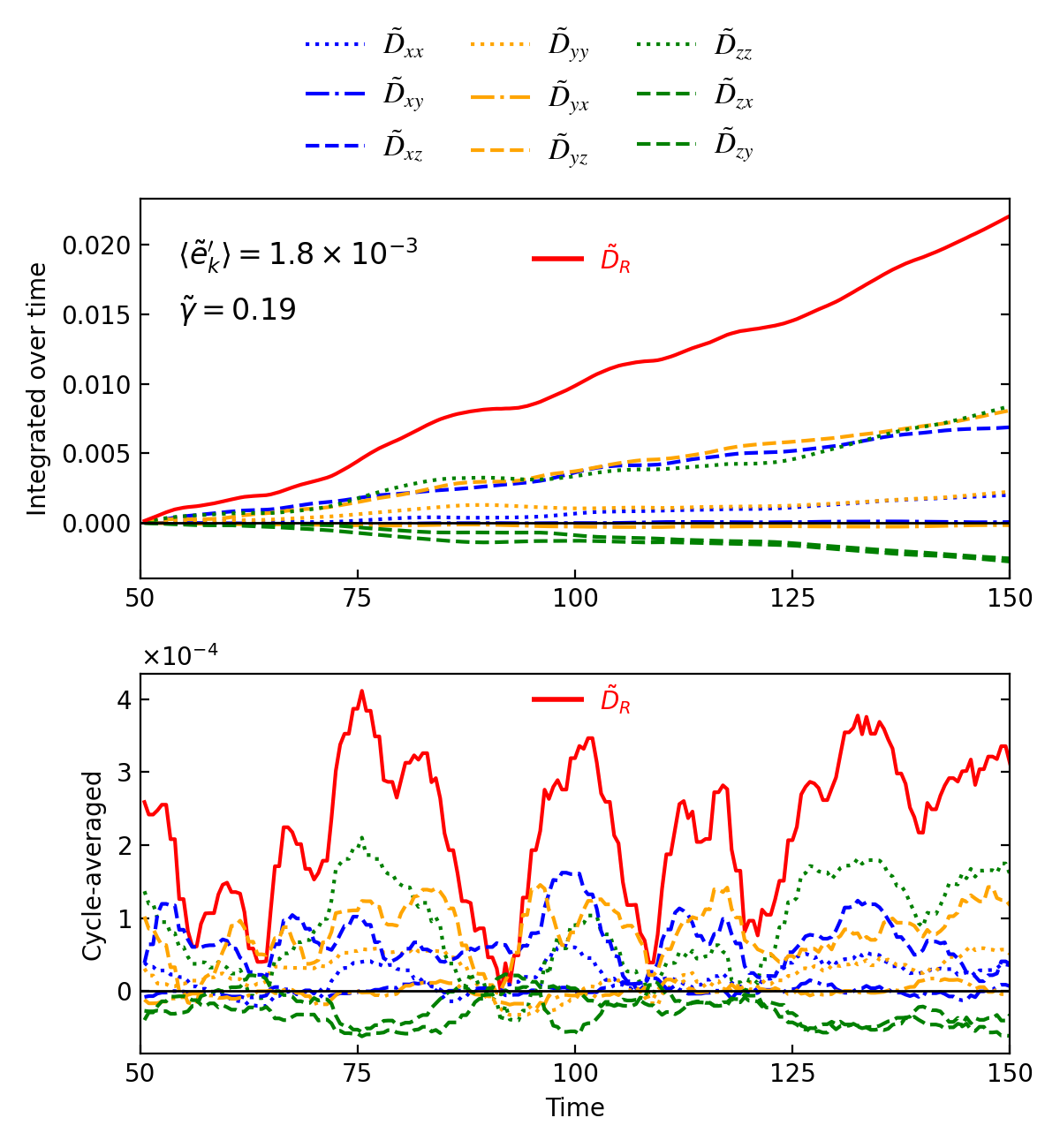}
\caption{
  Cumulative time integrals (upper panel) and cycle-averaged values 
(lower panel) of the transfer components for the potential-derived 
forcing $\tilde{\bf f}_4 = -\nabla \Psi$ at $\tilde{t}_{\rm osc}=0.5$, 
shown in the same format as Figure~\ref{fig:figure5}. 
The indicated value of $\langle \tilde{e}'_k \rangle$ denotes the time- and volume-averaged fluctuating kinetic energy over the interval displayed. 
As in the vertical-forcing case, the net positive transfer $\tilde{D}_R$ 
is dominated by $\tilde{D}_{zz}$, $\tilde{D}_{xz}$ and $\tilde{D}_{yz}$. 
However, these components show much stronger temporal fluctuations.
}
\label{fig:figure8}
\end{figure*}

Figure~\ref{fig:figure8} presents  the cumulative time integrals of the individual  transfer components, together with their cycle-averaged values.  As in the purely vertical forcing case, 
the  positive volume-averaged transfer $\tilde{D}_R$ is primarily associated with the components $\tilde{D}_{zz}$, $\tilde{D}_{xz}$ and $\tilde{D}_{yz}$,  with $\tilde{D}_{xx}$ and $\tilde{D}_{yy}$ providing smaller additional contributions. In contrast to the vertical-forcing case, however, all  components exhibit significantly stronger temporal variability. This enhanced variabili\-ty, together with the significant surface contribution in the mean-flow and fluctuation budgets, reflects the fact that net work is delivered at the boundary rather than in the fluid interior.

To test the robustness of these results, we have also performed simulations with $\tilde{t}_{\rm osc}=1$ and with alternative polynomial (rather than sinusoidal) choices of $\Psi(x,y)$.  

In the simulations shown, the maximum surface displacement is $0.04$, confirming that the small-deformation assumption (relative to $L_z$) is well satisfied.

\section{Origin of the Reynolds-stress correlations }
\label{sec:discussion}

The results presented in the previous section show that,  over the parameter range investigated here,  the volume-averaged transfer term $D_R$ is consistently positive, implying  a net transfer of kinetic energy from the fluctuations to the mean flow.  
This transfer arises from systematic
correlations between the fluctuating velocity field and gradients of the
mean flow. In this section, we discuss the physical origin of 
these correlations.


\subsection{Forced and secondary fluctuations}
\label{sec:secondaryfluc}

For the forcing defined in equation~(\ref{eq:forcing1}), only the vertical
velocity component is driven directly, generating a fluctua\-ting motion
$u'_z$. The horizontal velocity fluctuations $u'_x$ and $u'_y$ therefore
arise indirectly: the forced vertical motion, being subject to buoyancy and incompressibility constraints, is redirected in the same manner as the mean convective flow, generating secondary horizontal fluctuations.  As a result, correlations between the Reynolds stresses and the mean-flow velocity gradients naturally develop.
 Similar considera\-tions apply when
the forcing acts purely in the horizontal direction.
 More generally, even when the forcing includes both vertical and horizontal components, secondary fluctuations arise and establish correlations with the mean-flow gradients.

To clarify this mechanism, we decompose the fluctuating
velocity as
${\bf u}' = {\bf u}'_{\rm f} + {\bf u}'_{\rm s}$, where
${\bf u}'_{\rm f}$ denotes the direct response to the external forcing
and ${\bf u}'_{\rm s}$ represents secondary fluctuations.
Although
such a decomposition is not strictly valid in a  fully nonlinear
system, it provides a useful conceptual framework.

 For example,  a vertically oscillatory forced velocity ${u}'_{{\rm f},z} (x,y,z)$  induces horizontal velocity fluctuations ${u}'_{{\rm s},x}$ and ${u}'_{{\rm s},y}$ with amplitudes  comparable to that of  ${u}'_{{\rm f},z}$, through  the incompressibility  constraint and buoyancy. The Reynolds stress $ \left< u'_x u'_z \right> \approx  \left< {u}'_{{\rm s},x} {u}'_{{\rm f},z} \right>$ then couples
efficiently to the mean vertical shear $\partial V_x/\partial z$.

Similarly, the forced vertical oscillation
${u}'_{{\rm f},z}$ generates a secon\-dary vertical component 
${u}'_{{\rm s},z}$ which couples to the mean vertical shear
$\partial V_z/\partial z$. Together, these mecha\-nisms provide a natural explanation
for the robust Reynolds-stress correlations observed between the fluctua\-ting
velocity field and the large-scale mean flow: the secondary fluctuations, being subject to the same buoyancy and incompressibility constraints as  the mean flow, naturally inherit its spatial structure.

This interpretation is supported by the fact that the for\-cings used in the simulations produce larger ${u}'_{{\rm f},z}$ in the upper regions of the domain, where  $\partial V_z/ \partial z$ is predominantly negative (see Fig.~\ref{fig:figure1}).  Nevertheless, across all cases examined, we find $D_{zz}>0$, implying that ${u}'^2_{z}$ is systematically larger in regions where the mean shear  $\partial V_z/ \partial z$ is positive.   This counterintuitive spatial preference reflects the fact that the secondary fluctuations, being subject to the same buoyancy constraints as the mean flow, are preferentially amplified in regions of stronger convective instability, where $\partial V_z/ \partial z >0.$

This picture also explains why the computation of $D_R$ is comparatively insensitive to the details of the numerical scheme, in contrast to the external work $\left< {\bf f} \cdot {\bf u}' \right>$, as noted earlier.  The latter depends on small phase offsets between $ {\bf f} $ and $ {\bf u}'$, which arise from the damping of  ${\bf u}'$ and are therefore very sensitive to numerical errors.  By contrast, the volume-averaged transfer $D_R$ results from spatial correlations between velocity fluctuations and slowly varying mean-flow gradients, which are controlled by the large-scale convective dynamics and are consequently much more robust to discretisation and time-stepping choices.

It is important to note that the directly forced oscillations alone may produce a Reynolds-stress tensor that does {\em not} correlate with the mean-flow velocity gradients.  The energy transfer  is therefore controlled primarily by the  contribution from the secondary fluctuations.  This explains why the
detailed structure of the forcing has only a weak influence on the resulting
transfer rates, as shown in the previous section. To illustrate this point, one may write:
$$ \left< u'_i u'_j \right> \approx  \left< {u}'_{{\rm f},i} {u}'_{{\rm f},j} \right> +  \left< {u}'_{{\rm f},i} {u}'_{{\rm s},j} \right> + \left< {u}'_{{\rm s},i} {u}'_{{\rm f},j} \right> + \left< {u}'_{{\rm s},i} {u}'_{{\rm s},j} \right>.$$
The first (purely forced) term on the right-hand side is expected to  contribute little to $D_{ij}$, while the mixed terms involving both forced and secondary
fluctuations provide the dominant source of correlation with the mean-flow
velocity gradients.  Since the amplitudes of the
secondary fluctuations are comparable to those of the forced oscillations,
these correlations naturally yield Reynolds stresses of order $u_f'^2$.

This mechanism applies equally to potential-derived forcings, such as $\tilde{\bf f}_4 = -\nabla \Psi$. Although the net work done by the forcing is achieved exclusively through the non-zero  boundary term (see Section~\ref{sec:potential}), the scalar potential $\Psi$ induces oscillatory displacements of fluid elements throughout the fluid interior. 
These motions generate secondary fluctuations that develop the
same correlations with the mean flow as in the non-potential cases.
Consequently, Reynolds-stress correlations are established throughout the
domain and sustain a net transfer to the mean flow.
 With a rigid upper boundary, the net work done by the forcing vanishes, so
any Reynolds-stress correlations must cancel in the volume average and
cannot produce a systematic transfer. When a free surface is present, the
non-zero surface work provides a net energy input, allowing these
correlations to organise into a sustained positive transfer.

\subsection{How the convective flow produces the correlations}
\label{sec:origincorrel}

A detailed analysis of the simulations and description of the mechanisms responsible for all observed  correlations will be presented in a subsequent paper.  Here we outline the
physical processes underlying those Reynolds-stress components for which a
clear interpretation can be identified from the simulations.  To this end, it is helpful to refer to the snapshots in Figure~\ref{fig:figure1}, which  illustrate the typical velocity field and its vertical gradients.  The full velocity field $\tilde{\bf u}$ shown in these snapshots   is nearly identical to the mean-flow velocity $\tilde{\bf V}$ entering  the transfer terms $\tilde{D}_{ij}$.  \\

For the case of a free upper surface, for which the mean-flow
velocity gradients can develop without direct kinematic constraints imposed by the
boundary, we have found in all simulations that the volume-averaged $D_{zz}$, $D_{xz}$ and $D_{yz}$ are positive, whereas the volume-averaged $D_{zx}$ and $D_{zy}$ are negative.  


\subsubsection{$D_{zz} > 0$}

Consider a plume moving upwards ($V_z>0$).  A positive vertical fluctuation $u'_z>0$ displaces a fluid parcel upward faster than the mean flow into a cooler environment. As a result, it has not cooled as much (via diffusion or mixing) and  is therefore  more buoyant.  This increased buoyancy then accelerates the parcel further upward and reinforces the original positive $u'_z$.  By contrast, for a negative $u'_z < 0$, the parcel is displaced downward into a hotter environment,  reducing buoyancy, which opposes the fluctuation and tends to restore the parcel toward the mean velocity.

In  regions where $\partial V_z / \partial z >0$, the coupling between $u'_z$ and $\delta T$ is more efficient: the growth rate of the convective instability is higher. Thus, aligned fluctuations ($u'_z > 0$) grow faster, while misaligned ones ($u'_z < 0$) are damped more effectively. In contrast, where $\partial V_z / \partial z < 0$, the buoyant drive is weaker, so the reinforcement/opposition is less pronounced. The same argument applies symmetrically to downward-moving plumes ($V_z < 0$), where negative fluctuations are preferentially reinforced in accelerating regions ($\partial V_z / \partial z > 0$).

This selective amplification in both rising and descending regions yields a robust positive   volume-averaged $D_{zz}= \left< u_z'^2 \right> \left( \partial V_z/\partial z \right)$. 

Figure~\ref{fig:figure5} shows that a larger volume-averaged values of $D_{zz}$ are obtained for smaller  oscillation periods ${t}_{\rm osc}$. Although decreasing ${t}_{\rm osc}$ at fixed fluctuation amplitude reduces the vertical displacement per cycle ($\delta z \sim u'_z {t}_{\rm osc}/2$), which would tend to weaken the buoyancy contrast and thus the selective reinforcement of $u'_z$, a competing effect dominates.   As ${t}_{\rm osc}$ decreases, the dimensionless dissipation length scale $\lambda_{\rm dis}$ increases (from $\sim$0.07 to $\sim$0.2 in the simulations) because nonlinear interactions have less time to cascade energy to small scales. Consequently, the dimensionless diffusive timescale $\tau_{\rm diff} \sim \lambda_{\rm dis}^2 / \tilde{\kappa}$ grows and more readily exceeds $\tilde{t}_{\rm osc}$, preserving buoyancy perturbations more effectively (closer to adiabatic conditions). This enhances the reinforcement mechanism despite the smaller displacements, resulting in stronger correlations and larger volume-averaged $D_{zz}$ at lower ${t}_{\rm osc}$.
 
\subsubsection{ $D_{zx}$ and $D_{zy}<0$}   

We focus on the term $D_{zx}$, with a similar argument applying to  $D_{zy}$.  Consider a rising plume  in a region where $\partial V_z/\partial x <0$, meaning the mean buoyancy $\delta T$ decreases  in the positive $x$-direction.  A positive horizontal fluctuation $u'_x>0$ displaces a fluid parcel towards  lower mean buoyancy.  Since the parcel approximately  conserves its temperature perturbation  over this short displacement, it becomes relatively  hotter (more buoyant) than its new surroundings,  accelerating upwards and acquiring a positive vertical fluctuation $u'_z>0$.  Conversely,  a negative horizontal fluctuation $u'_x<0$  displaces the parcel towards  higher mean buoyancy, making it relatively cooler (less buoyant) and thus generating  $u'_z<0$.  In both cases, $u'_x$ and the induced $u'_z$ have the same sign, yielding  $D_{zx}=\left<  u'_x  u'_z \right>  \left( \partial V_z/\partial x \right) <0$.   An analogous argument applies where $\partial V_z/\partial x >0$: displacements now induce opposite signs in $u'_x$ and $u'_z$,  yielding  again $D_{zx}<0$. The same logic holds for displacements along the $y$-direction.  

\subsubsection{  $D_{xz}$ and $D_{yz} >0 $}  

The physics mechanism responsible for the positive volume-averaged values of $D_{xz}$ and $D_{yz}$ is less clear than for the vertical contributions discussed above, and the interpretation we offer here remains speculative.  A more thorough analysis will be carried out separately. Consider a rising plume in a region where  $\partial V_x/\partial z >0$, corresponding to mean-flow vorticity $\boldsymbol{\omega} = \nab \times {\bf V} = \left( \partial V_x/\partial z \right) \hat{\bf y}$.  
A positive vertical fluctuation $u'_z (z) >0$ displaces a fluid parcel upward into layers where $V_x$ is larger.  One might therefore expect pressure and viscous stresses to adjust the parcel velocity toward that of the surrounding flow, leading to a vanishing horizontal fluctuation $u'_x$.   However, the fact that $D_{xz}>0$ indicates that the parcel acquires a positive horizontal fluctuation $u'_x>0$, so that it locally overtakes the mean horizontal flow.  One possible explanation is that the parcel experiences a lift-like acceleration associated with the background vorticity, analogous to the lift force acting on a body in a rotational flow \citep{Auton1987}.  To illustrate this idea, suppose the fluid parcel were replaced by a small bubble at rest relative to the laboratory frame. In that case, the undisturbed flow would move past the bubble with a relative velocity $\mathbf{U}_{\rm rel}=-u'_z\,\hat{\mathbf z}$, and the background vorticity would generate a lift force proportional to $\mathbf{U}_{\rm rel} \times \boldsymbol{\omega}$, directed in the positive $x$-direction. Such a mechanism would accelerate the bubble horizontally beyond the local mean-flow speed.
While a fluid parcel is not a rigid body and does not experience a literal Magnus force, this analogy suggests that interactions between vertical slip motions and mean-flow vorticity can bias the horizontal response of rising and sinking plumes, leading to positive correlations$\langle u'_x u'_z\rangle$ (and similarly $\langle u'_y u'_z\rangle$) in regions where $\partial V_x/\partial z>0$.

\subsubsection{Viscous and thermal constraints on the transfer mechanism}

These mechanisms are consistent with the results presented in Section~\ref{sec:Rayleighnumber}, where the effect of varying the Rayleigh number was examined.  Increasing viscous diffusion (i.e. decreasing $Ra$) reduces the relative contributions of $D_{xz}$ and $D_{yz}$ to the net transfer, because horizontal motions across a convective length scale are increasingly damped by viscosity.  By contrast, the $D_{zz}$ contribution, which is primarily buoyancy-driven, is comparatively less affected.  The criterion~(\ref{eq:Reynoldsfluc}), involving the Reynolds number of the fluctuations, therefore reflects the conditions required to establish correlations in the $D_{xz}$ and $D_{yz}$ terms, rather than in $D_{zz}$.  

The mechanisms described above also rely on fluid parcels retaining their temperature perturbations as they are displaced over a convective length scale.  This requires thermal diffusion to be sufficiently weak over the displacement timescale. We therefore expect an additional constraint, analo\-gous to equation~(\ref{eq:Reynoldsfluc}), involving the fluctuating P\'eclet number: 
\begin{equation}
Pe' \equiv \frac{u’ \lambda_{\rm conv}}{\kappa} ,
\label{eq:Pecletfluc}
\end{equation}
which must be of order unity or larger for the buoyancy-mediated correlations to be sustained. This will be investigated in a subsequent paper. 

\subsubsection{The role of spatial coherence in the fluctuation–mean flow coupling}

 The mechanisms described above rely on fluid parcels being displaced by the fluctuating motions into regions where local mean-flow properties, such as buoyancy or velocity, differ from those at their original location.  These displacements are  small, of order $u' t_{\rm osc}$, where $u'$ is a characteristic amplitude of the fluctuating velocity.  Nevertheless, because the fluctuations are coherent over large spatial scales comparable to convective plume widths, their cumulative effect becomes significant when integrated over the volume. This spatial coherence allows the resulting accelerations, whether buoyancy-driven or arising from other interactions with the mean flow, to act systematically in phase over extended regions.  
  As a result,   robust volume-averaged correlations develop between $\langle u'_i u'_j \rangle$ and the mean-flow gradients $\partial V_i / \partial x_j$,  yielding finite contributions to the transfer terms $D_{ij}$.

\subsubsection{Spatial coherence dominates over timescale matching}

It is often assumed that correlations between oscillatory fluctuations and the mean flow require a matching of timescales between the forcing and the convective dynamics. Our results show that this is not the case. Instead, the emergence of a positive $D_R$ reflects a match in spatial, rather than temporal, scales between the oscillatory motions and the convective structures.  In particular, no resonance between the oscillation period and convective turnover times is required.

The oscillation period is largely irrelevant, provided that fluid elements can be displaced back and forth coherently over an oscillation period without viscous or thermal diffusion overwhelming advection. This condition is satisfied as long as the fluctuating Reynolds and Péclet numbers meet the criteria given by equations~(\ref{eq:Reynoldsfluc}) and~(\ref{eq:Pecletfluc}).  By contrast, matching the relevant length scales is essential: spatial coherence over a convective length scale is required for the correlations to build up and yield a net energy transfer, as described above.

This perspective represents a significant shift in our understanding of tidal dissipation, emphasizing the primacy of spatial coherence over timescale matching in mediating the interaction between oscillatory forcing and convective flows.

\subsubsection{Effect of the boundary condition on the correlations}

The free upper surface plays two conceptually distinct roles in the simulations, which we distinguish here. The first is energetic: as shown by equation~(\ref{eq:workpotential}), a forcing derived from a scalar potential can perform net work on the flow only through a non-zero surface term, which requires a free surface. This is a mathematical result independent of the details of the flow. The second role is dynamical: the free surface permits large-scale vertical plume structures and mean-flow gradients that differ qualitatively from those found with a rigid lid, and these differences affect the Reynolds-stress correlations, 
as discussed in Section~\ref{sec:rigid}.

With a rigid boundary,   $D_{xx}$  becomes positive when the for\-cing includes a horizontal component, while $D_{zz}$ no longer contributes to the transfer.  The impermeability constraint imposed by the rigid lid redirects part of the vertically forced response into strong, strain-dominated horizontal motions near the upper boundary.  As a result, regions of enhanced $u_x'^2$  pre\-ferentially coincide with positive horizontal strain.
The signs of $D_{zx}$ and $D_{xz}$ are also reversed relative to the free-surface case. This indicates that, in
the presence of a rigid boun\-dary, the pressure forces required to enforce
impenetrability substantially  modify the mean-flow gradients and constrain the
fluctuation dynamics.  Consequently,  the mechanisms operating in the free-surface case
are no longer dominant. In this sense, the rigid-surface configuration
represents a qualitatively distinct regime, in which coupling between
the fluctuations and the mean flow is governed  primarily by boun\-dary constraints
and pressure adjustments rather than by the intrinsic structure of the convective flow.  

A more complete physical explanation of why the rigid boundary suppresses or modifies the correlations would require a detailed analysis of the mean-flow structure and plume morphology in the two configurations. This goes beyond what we can establish from the present simulations and we flag it as an important question for future work.

\section{Summary and discussion}
\label{sec:summary}

\subsection{Summary}

The simulations presented in this paper show that, when the upper surface is free and the fluctuating Reynolds number exceeds  a modest threshold,  
 the oscillations transfer kinetic energy systematically to the mean flow. The volume-averaged transfer rate is 
$D_R \sim  u'^2 \, t_{\rm conv}^{-1}$,
where $u'$ denotes a characteristic amplitude of the fluctuating velocity.
This corresponds to an effective dimensionless damping rate $\tilde{\gamma} \simeq 0.1$,  essentially independent of the oscillation period over the range explored. This behaviour reflects  robust correlations between the Reynolds stresses of the fluctuations and the gradients of the large-scale convective flow.  

A decomposition of the transfer by Reynolds-stress component shows that, for a free upper surface, the dominant positive contributions typically involve vertical velocity fluctuations interacting with vertical gradients of the mean flow. Contributions associated with horizontal shear tend to oppose this transfer and partially cancel the buoyancy-driven term, such that the net transfer is well approximated by the mixed components involving vertical motions. These correlations arise because the oscillations coherently displace fluid parcels over large spatial regions comparable to plume widths. The displaced parcels are then systematically redirected by buoyancy and incompressibility, producing sustained, volume-filling correlations.  

Importantly, potential-derived forcings that mimic realistic tidal perturbations yield the same robust positive transfer rates and dominance of vertical correlations. This demonstrates that the mechanism remains effective even when the net energy input is localised at the surface.

Viscous and thermal diffusion place clear constraints on this mechanism. Positive transfer is observed only when the fluctuating Reynolds number satisfies $Re' \gtrsim \mathcal{O}(1\!-\!10)$, en\-suring that advective displacements across convective length scales are not erased by viscous damping.  We expect an analogous constraint involving the fluctuating P\'eclet number, which reflects the need for fluid parcels to retain their temperature perturbations while being displaced. 

Replacing the free upper surface with a rigid boundary has two distinct consequences. First, for  forcings derived from a potential, a rigid boundary prevents the forcing from performing net work on the flow, as shown by equation~(\ref{eq:workpotential}). Se\-cond, and independently, the rigid boundary alters the large-scale structure of the mean flow and the associated Reynolds-stress correlations, suppressing the dominant transfer terms present in the free-surface case. The physical origin of this second effect remains unclear, and distinguishing the relative contributions of energetic consistency, plume morphology and mean-flow gradients is left for future work.
In summary, the rigid-surface configuration  represents a distinct regime that is not representative of stellar or planetary convective envelopes.

\subsection{Discussion}
\label{sec:discussion2}

\citet{Barker2021} argued that the volume-averaged transfer term $D_R$ does not contribute to the damping of fast oscillations in convective flows. This conclusion was based on a combination of analytical arguments and numerical simulations.  As already noted by \citet{Terquem2023}, their analy\-tical argument is flawed, owing to a misidentification of the term responsible for energy exchange between the oscillatory fluctuations and the mean flow. In addition, their numerical simulations impose a rigid upper boundary, a configuration that we have shown suppresses the correlations required to produce a net energy transfer from the fluctuations to the mean flow.
 More fundamentally, their simulations prescribe the oscillatory velocity field rather than allowing it to emerge self-consistently as the response to an external forcing. Speci\-fically, the Navier–Stokes equation solved for the mean flow contains source terms involving an irrotational tidal velo\-city derived from a prescribed potential. As demonstrated in this paper, the correlations responsible for a positive $D_R$ arise because fluid displacements driven by the external forcing generate secondary oscillations that subsequently respond to buoyancy and incompressibility constraints.
Prescribing an oscillatory velocity field directly in the advection terms therefore interferes with the self-consistent development of the secon\-dary responses and the associated correlations.  Capturing the correct energy exchange requires allowing the flow to adjust dynamically to the forcing, rather than imposing the oscillatory motion a priori. This self-consistent adjustment is essential for the emergence of the correlations underlying the transfer mechanism identified here.

A key result of this study is that $D_R$ is independent of the forcing period over the range explored.   This period-independence is not merely an empirical observation. Rather, it follows from the physical argument developed in Section~\ref{sec:origincorrel}: the correlations responsible for $D_R$ arise from spatial coherence between the oscillatory displacements and the convective structures, and do not require any mat\-ching of timescales between the forcing and the convection. The scaling $D_R \sim u'^2 t_{\rm conv}^{-1}$ contains no explicit dependence on $t_{\rm osc}$, and period-independence is therefore expected to hold throughout the regime $t_{\rm osc} \ll t_{\rm conv}$, provided that  the fluctuating Reynolds and P\'eclet number conditions given by equations~(\ref{eq:Reynoldsfluc}) and~(\ref{eq:Pecletfluc}) are satisfied. That said, our numerical simulations span only a factor of ten in $t_{\rm osc}$, and direct verification over a broader range would be valuable.

Although our simulations were performed at Rayleigh and Prandtl numbers far from those relevant to stellar and plane\-tary convection, they nevertheless isolate physical mechanisms that are expected to persist in more extreme regimes. In particular, the energy transfer quantified by $D_R$ arises from correlations between oscillatory displacements and the large-scale convective flow, which depend primarily on spatial coherence and on the ability of the fluctuations to advect fluid across mean-flow gradients. These conditions are controlled by the fluctuating Reynolds and P\'eclet numbers, rather than by the absolute values of $Ra$ and $Pr$, and are therefore expected to remain satisfied in astrophysical environments where $Re'$ and $Pe'$ are extremely large.
In the outer convective envelope of a solar-type star, above roughly 0.8~R$_\odot$, the region most relevant for tidal interactions with a companion, $Ra \sim 10^{20}$--$10^{22}$ and $Pr \sim 10^{-6}$--$10^{-4}$ \citep{Schumacher2020}.   These values correspond to thermal diffusivities $\kappa \sim 1$--$500$~m$^2$~s$^{-1}$ and kinematic viscosities $\nu \sim $ a few times $10^{-4}$~m$^2$~s$^{-1}$.   The characteristic convective length scale in this region is $\lambda_{\rm conv} \sim 0.08$~R$_\odot$ \citep{Terquem2021}.
The tidal velocity amplitude excited by a 1~M$_\odot$ companion is typi\-cally in the range 0.01--1~m~s$^{-1}$ for orbital periods between 4 and 12 days (\citealt{Terquem1998}, \citealt{Terquem2021}).  These va\-lues imply fluctuating Reynolds and P\'eclet numbers  $Re' > 10^9$ and $Pe' > 10^3$, well above the thresholds required for the establishment of the correlations between the fluctuations and the mean flow. In this regime of extremely large fluctuating Reynolds numbers, we  expect the horizontal–vertical shear terms $D_{xz}$ and $D_{yz}$ to dominate tidal damping. 
At the same time, important uncertainties remain in extrapolating these results quantitatively to astrophysical conditions. The structure of convection at $Ra \sim 10^{20}$--$10^{22}$ is significantly more turbulent and intermittent than in the present simulations, and may modify the detailed distribution of velocity gradients and the resulting correlations. In addition, the influence of very low Prandtl number, strong stratification, rotation and magnetic fields, all absent from our model, may affect both the magnitude and spatial structure of the energy transfer. While these effects are not expected to suppress the mechanism itself, they may either enhance or reduce its efficiency.


The results presented in this paper validate the formalism proposed by Terquem (2021, 2023),
who demonstrated from first principles that the traditional turbulent-viscosity description breaks down when the external forcing varies on a  timescale much shorter than the convective timescale. 
For fast tides, the relevant Reynolds stresses are  set by correlations involving the tidal velocity fluctuations and gradients of the convective flow, rather than by correlations involving convective velocities and gradients of the tidal flow.

Adopting this formalism, \citet{TerquemMartin2021} explored the consequences of equilibrium-tide dissipation by assuming an energy transfer rate of the form investigated here. Under this assumption, they showed that dissipation of the equilibrium tide alone can account for the observed circularisation periods of solar-type binaries.  By incorpora\-ting, for the first time, the full time evolution of stellar structure, they found that tidal dissipation is efficient both before and after the main sequence, while remaining weak during the main-sequence phase, in significantly improved agreement with observations compared to previous theories.  Using the same framework, \citet{Terquem2023} derived an average phase lag between the equilibrium tide and the tidal potential due to moons, obtained by integrating over the convective envelope of giant planets, in good agreement with observational constraints for Jupiter and several of Saturn’s moons.

The present results establish the existence and basic scaling of the mechanism by which fast tides transfer their energy to the mean convective flow.  Further work is required to explore a broader region of parameter space and to develop a more detailed physical understanding of the correlations underlying this transfer. In particular, extending this study to a wider range of Rayleigh and Prandtl numbers, and to for\-cing prescriptions that more closely reflect tidal potentials in stratified convective envelopes, will help clarify the  origin of these correlations  and their dependence on viscosity, thermal diffusion and geometry.

Taken together, these results represent a substantial advance in our understanding of tidal dissipation, with broad implications for the evolution of stellar binaries, planetary systems and the dynamics of forced turbulent flows more generally.

\section*{Acknowledgements}

AB is supported by a PhD studentship from the Science and Technology Facilities Council (STFC) under grant ST/Y509474/1.  EM was supported by the Summer Undergraduate Research Opportunities Programme (UROP) hosted by the Rudolf Peierls Centre for Theoretical Physics.   CT acknowledges the use of ChatGPT and Grok in developing the post-processing codes for data produced by the Dedalus simulations. CT also thanks  the Isaac Newton Institute for Mathematical Sciences, Cambridge, for its support and hospitality during the programme {\em Anti-diffusive dynamics: from sub-cellular to astrophysical scales}, where part of this work was undertaken. This programme was supported by EPSRC grant EP/R014604/1.  CT is grateful to Sacha Brun, Stephan Fauve and Antoine Strugarek for illuminating discussions on convection, to Geoff Vasil for advice on Dedalus and convection more broadly, and to Steven Balbus for sharing his invaluable expertise in fluid dynamics and for his unfailing optimism regarding the outcome of this work.  Finally, we thank the referee for a thoughtful and insightful report and for suggestions that have improved the paper.

\section*{Data availability}

No new data were generated or analysed in support of this research.










\appendix

\section{Derivations of free-surface boundary conditions}
\label{appendixA}

We consider a free surface $z= H + \eta \left( x, y \right) $ which radiates as a blackbody at temperature $T_2$.  

\subsection{Kinematic boundary condition}

\noindent At the surface  $z=H+\eta$, the kinematic boundary condition is $u_z = {\rm d} \eta / {\rm d} t$.  Defining dimensionless quantities $\tilde{\eta}=\eta/d$ and $L_z=H/d$, this becomes:

\begin{equation}
\tilde{u}_z  =   \tilde{\eta}_{\tilde{t}}
  +  \tilde{u}_x   \tilde{\eta}_{\tilde{x}} + \tilde{u}_y    \tilde{\eta}_{\tilde{y}}  ,
\end{equation}

\noindent where all velocity components are evaluated at $\tilde{z}=L_z + \tilde{\eta}$ and we define:

\begin{equation}
 \tilde{\eta}_{\tilde{t}}  \equiv
  \frac{\partial \tilde{\eta}}{\partial \tilde{t}} , \quad   \tilde{\eta}_{\tilde{x}} \equiv   \frac{\partial  \tilde{\eta} }{\partial \tilde{x}}  , \quad    \tilde{\eta}_{\tilde{y}} \equiv   \frac{\partial   \tilde{\eta}}{\partial \tilde{y}} .
  \label{eq:etaderiv}
\end{equation}

Assuming small deformation, i.e.  $ \left| \tilde{\eta} \right| \ll L_z$, we expand to first order:

\begin{equation}
\tilde{u}_z + \tilde{\eta} \frac{\partial \tilde{u}_z}{\partial \tilde{z}} =   \tilde{\eta}_{\tilde{t}}
  +  \tilde{u}_x   \tilde{\eta}_{\tilde{x}} + \tilde{u}_y   \tilde{\eta}_{\tilde{y}}  ,
  \label{eq:etasurf}
\end{equation}

\noindent  evaluated at $\tilde{z}=L_z$.

\subsection{Continuity of the tangential stress}

At the surface  $z = H+\eta \left( x, y \right)$, we define tangent vectors:

\begin{equation}
\hat{\bf t}^{(x)} = \frac{1}{\sqrt{1+ \eta_{x}^2}} 
\begin{pmatrix}
1 \\
0 \\
\eta_x
\end{pmatrix} , \quad
\hat{\bf t}^{(y)} = \frac{1}{\sqrt{1+ \eta_{y}^2}} 
\begin{pmatrix}
0 \\
1 \\
\eta_y
\end{pmatrix},
\end{equation}

\noindent  and the surface normal:

\begin{equation}
\hat{\bf n} = \frac{1}{\sqrt{1+ \eta_x^2 + \eta_{y}^2}} 
\begin{pmatrix}
- \eta_x \\
- \eta_y \\
1
\end{pmatrix},
\end{equation}

\noindent where $\eta_x = \partial \eta / \partial x = \tilde{\eta}_{\tilde{x}}$ and $\eta_y= \partial \eta / \partial y = \tilde{\eta}_{\tilde{y}}$.

\noindent 
The viscous stress is  $T_i=\sigma_{ij} n_j$, where $\sigma_{ij}$ are the components of the stress tensor and $n_j$ are the components of $\hat{\bf n}$.  For an incompressible Newtonian fluid:

\begin{equation}
\sigma_{ij}=\sigma_{ji}=\rho \nu \left(  \frac{\partial u_i}{\partial x_j} +   \frac{\partial u_j}{\partial x_i}   \right),
\end{equation}

\noindent with $x_1=x$, $x_2=y$ and $x_3=z$.  

Continuity of tangential stress at the surface requires ${\bf T} \cdot \hat{\bf t}^{(x)} =0$, which gives:

\begin{equation}
\eta_x \left( \sigma_{zz} - \sigma_{xx} \right) + \left( 1 - \eta_x^2 \right) \sigma_{xz} 
-\eta_y \left(  \sigma_{xy} + \eta_x \sigma_{zy} \right) = 0,
\end{equation}

\noindent where the components of the stress tensor are evaluated at $z=H+ \eta$.  In dimensionless form and to first order in  $ \left| \tilde{\eta} \right| / L_z$, this becomes:

\begin{equation}
\tilde{\eta}_{\tilde{x}} \left( \tilde{\sigma}_{zz} - \tilde{\sigma}_{xx} \right) +  \tilde{\sigma}_{xz} + \tilde{\eta} \frac{\partial \tilde{\sigma}_{xz}}{\partial \tilde{z}}
-\tilde{\eta}_{\tilde{y}}  \tilde{\sigma}_{xy}  = 0,
\label{eq:stressx}
\end{equation}

\noindent  evaluated at $\tilde{z}=L_z$.

\noindent Similarly, the condition ${\bf T} \cdot \hat{\bf t}^{(y)} =0$ yields:

\begin{equation}
\tilde{\eta}_{\tilde{y}} \left( \tilde{\sigma}_{zz} - \tilde{\sigma}_{yy} \right) +  \tilde{\sigma}_{yz} + \tilde{\eta} \frac{\partial \tilde{\sigma}_{yz}}{\partial \tilde{z}}
-\tilde{\eta}_{\tilde{x}}  \tilde{\sigma}_{xy}  = 0,
\label{eq:stressy}
\end{equation}

\noindent also at $\tilde{z}=L_z$.

\subsection{Continuity of the normal stress}

Let  $P_{\rm atm}$ be the  atmospheric pressure above the fluid.  At $z=H+\eta$, the normal stress balance is:

\begin{equation}
{\bf T} \cdot \hat{\bf n} - P = - P_{\rm atm} ,
\label{eq:stressn1}
\end{equation}

\noindent We write $P=P_0 + \delta P$, with $P_0$ satisfying the hydrostatic equilibrium equation~(\ref{eq:NSeq}).  This yields $P_0=-\rho_0gz + C$, where the constant $C$ is determined by the  boundary condition $P_0=P_{\rm atm}$ at $z=H$.  Substituting in equation~(\ref{eq:stressn1}) yields:

\begin{equation}
{\bf T} \cdot \hat{\bf n} +   \rho_0 g  \eta   - \delta P = 0 ,
\label{eq:stressn2}
\end{equation}

\noindent evaluated at $z=H+\eta$.  In dimensionless form:

\begin{equation}
\tilde{{\bf T}} \cdot \hat{\bf n} + \frac{\tilde{\eta}}{\alpha \Delta T} - \tilde{P} = 0,
\end{equation}

\noindent evaluated at $\tilde{z}=L_z + \tilde{\eta}$, where $\tilde{T}_i = \tilde{\sigma}_{ij} n_j$.  Expanding this using the stress components:

\begin{equation}
\begin{split}
\frac{1}{\sqrt{1+ \tilde{\eta}_{\tilde{x}}^2 + \tilde{\eta}_{\tilde{y}}^2}} 
\Big(  
& \tilde{\eta}_{\tilde{x}}^2 \tilde{\sigma}_{xx} 
+ \tilde{\eta}_{\tilde{y}}^2 \tilde{\sigma}_{yy} 
+ \tilde{\sigma}_{zz} 
+ 2 \tilde{\eta}_{\tilde{x}} \tilde{\eta}_{\tilde{y}} \tilde{\sigma}_{xy} \\
& -2  \tilde{\eta}_{\tilde{x}} \tilde{\sigma}_{xz} 
-2  \tilde{\eta}_{\tilde{y}} \tilde{\sigma}_{yz}
\Big) 
+ \frac{\tilde{\eta}}{\alpha \Delta T} - \tilde{P} = 0 ,
\end{split}
\end{equation}

\noindent  To first order in $ \left| \tilde{\eta} \right| / L_z$, this reduces to:

\begin{equation}
  \tilde{\sigma}_{zz}  + \tilde{\eta} \frac{\partial  \tilde{\sigma}_{zz} }{ \partial \tilde{z}}
-2  \tilde{\eta}_{\tilde{x}} \tilde{\sigma}_{xz} 
-2  \tilde{\eta}_{\tilde{y}} \tilde{\sigma}_{yz}
 + \frac{\tilde{\eta}}{\alpha \Delta T} - \tilde{P} - \tilde{\eta} \frac{\partial \tilde{P}}{\partial \tilde{z}} = 0,
\label{eq:stressnorm}
\end{equation}

\noindent evaluated at $\tilde{z}=L_z$.

\subsection{Radiative boundary condition}

At the surface, the radiative flux is given by:

\begin{equation}
-k \nab T \cdot {\bf n} = \epsilon \sigma \left( T^4 - T_{\rm atm}^4 \right),
\end{equation}

\noindent where $T$ is evaluated at $z=H+ \eta$,  $T_{\rm atm}$ is the atmospheric temperature  above the fluid, $\sigma$ is the Stefan-Boltzmann constant, $\epsilon$ is the emissivity and $k$ is the thermal conductivity.  We take 
$T_{\rm atm}=0$ and $\epsilon=1$, corresponding to a blackbody radiating into a vacuum.  Using $T=T_0 + \delta T$ and $\nab T_0  = -  \left( \Delta T / H \right) \, \hat{\bf z}$, the boundary condition becomes:

\begin{equation}
\frac{-k}{\sqrt{1+ \eta_x^2 + \eta_{y}^2}} \left(   
-\eta_x \frac{\partial \delta T }{\partial x} -\eta_y \frac{\partial \delta T}{\partial y} + \frac{\partial \delta T }{\partial z }  - \frac{\Delta T}{H}
\right)  =  \sigma  T^4 ,
\label{eq:radsurf1}
\end{equation}

\noindent evaluated at  $z=H+ \eta$.  To first order in $\left| \eta \right| / H$, we expand:

\begin{equation}
T^4 \left( H + \eta \right) = T^4 \left( H \right) + 4 \eta T^3  \left( H \right) \frac{\partial T}{\partial z} \left( H \right) .
\end{equation}

\noindent With $T \left( H \right) =T_2 + \delta T \left( H \right) $, where $T_2 \equiv T_0 \left( H \right)$, this becomes:

\begin{equation}
\begin{split}
 T^4 & \left( H  + \eta \right) =  T_2^4 \left( 1 +  \frac{\delta T \left( H \right)}{T_2} \right)^4 \\
& + 
4 \eta T_2^3 \left( 1 +  \frac{\delta T \left( H \right)}{T_2} \right)^3 \left( - \frac{\Delta T}{H} +  \frac{\partial \delta T }{\partial z }  \left( H \right) \right) .
\end{split}
\end{equation}

\noindent At equilibrium, $-k {\rm d} T_0 / {\rm d} z = k \Delta T / H = F_0$, and the surface temperature $T_2$ satisfies $\sigma T_2^4 = F_0$.  Substituting these relations yields:

\begin{equation}
 \sigma T^4 \left( H + \eta \right) =  F_0 \left[ 
 \left( 1 +  \frac{\tilde{b} }{\tilde{T}_2} \right)^4 - 
4 \tilde{\eta} \frac{1}{\tilde{T}_2} \left( 1 +  \frac{\tilde{b} }{\tilde{T}_2} \right)^3 \left( 1 -  \frac{\partial \tilde{b} }{\partial \tilde{z} }   \right) \right] ,
\label{eq:lhsrad}
\end{equation}

\noindent where $\tilde{b}$ and its derivative are evaluated at $\tilde{z}=L_z$ and we define $\tilde{T}_2 = T_2 / \Delta T$.

Finally, substituting into  equation~(\ref{eq:radsurf1}), we obtain, to 
 first order in $\left| \eta \right| / H$:

\begin{equation}
\begin{split}
 1+ \tilde{\eta}_{\tilde{x}}  \frac{\partial \tilde{b} }{\partial \tilde{x} } 
& + \tilde{\eta}_{\tilde{y}}  \frac{\partial \tilde{b} }{\partial \tilde{y} } 
 -  \frac{\partial \tilde{b} }{\partial \tilde{z}}  - \tilde{\eta} \frac{\partial^2 \tilde{b} }{\partial \tilde{z}^2} = 
  \\
& \left(1+ \frac{\tilde{b} }{\tilde{T}_2} \right)^4 - 
4 \tilde{\eta} \frac{1}{\tilde{T}_2} \left( 1 +  \frac{\tilde{b} }{\tilde{T}_2} \right)^3 \left( 1 -  \frac{\partial \tilde{b} }{\partial \tilde{z} }   \right),
\label{eq:surfrad}
\end{split}
\end{equation}

\noindent evaluated at $\tilde{z}=L_z$.

\bsp	
\label{lastpage}
\end{document}